\documentclass[preprint,12pt]{elsarticle}

\usepackage{amssymb}
\usepackage{hyperref}

\usepackage{lineno}

\newcounter{bla}

\journal{Computer Physics Communications}

\begin{document}

\begin{frontmatter}

\title{Simple: A pedagogic tool for nuclear reactions and particle interactions with matter}

\author[a]{Deepak Samuel\corref{author}}

\cortext[author] {Corresponding author.\\\textit{E-mail address:} deepaksamuel@cuk.ac.in}
\address[a]{Department of Physics, Central University of Karnataka, India 585367.}

\begin{abstract}
Nuclear and particle physics are the core components of most undergraduate and postgraduate physics courses worldwide. While few fundamental concepts like particle counting and detector characterisation are taught in tandem with laboratory sessions, advanced techniques like cross-section measurement, track reconstruction, etc., are usually not taught extensively. At the same time, modern particle physics experiments require students to have advanced skills in statistical treatment of data, which cannot be covered in detail in the regular coursework. A software named Simple has been developed to simulate particle physics experiments, based on which, skills necessary for particle physics studies can be easily imparted. Simple is based on Geant4 libraries, widely used by the high energy physics community to simulate particle interactions with matter. Unlike Geant4, Simple is designed to be a standalone application requiring no prior programming knowledge to run a simulation and the time consuming compilation step is also avoided. In this paper, the features of this software along with few sample simulations are presented. Though the emphasis of this paper is on the use of Simple as a pedagogical aid, this application can also be used as a general purpose simulation framework for particle physics analyses.
\end{abstract}


\end{frontmatter}
\newpage

{\bf PROGRAM SUMMARY}

\begin{small}
\noindent
{\em Program Title: Simple }                                        \\
{\em Licensing provisions(please choose one): LGPL}                  \\
{\em Programming language: C++}                                   \\
{\em Supplementary material: Demonstration video}                                 \\
{\em Journal reference of previous version:NA}                  \\
{\em Does the new version supersede the previous version?:NA}   \\
{\em Reasons for the new version:NA}\\
{\em Summary of revisions:NA}*\\
{\em Nature of problem(approx. 50-250 words):}\\
  The use of Geant4 as an aid for teaching nuclear physics is currently limited owing to the steep learning-curve which demands advanced knowledge in software.\\
{\em Solution method(approx. 50-250 words):}\\
 A front-end with fast access to key elements was designed to help the user simulate an experiment and analyse the data in a few clicks. The user is not required to compile Geant4 on the system as the programme is distributed as an standalone application.\\
{\em Additional comments including Restrictions and Unusual features (approx. 50-250 words):}\\
 The standalone application exists for Ubuntu 19.10. This needs to be created for other operating system.
   \\
\end{small}


\section{Introduction} 
Monte-carlo simulations have played a pivotal role in many nuclear and particle physics experiments. Geant4 is a C++ library used to simulate particle interactions with matter, which has become the mainstay in many particle physics experiments and has even found applications in other fields such as space science and medical physics \cite{agostinelli2003geant4}. Typical applications include studying detector response, bench-marking and validation of algorithms and track visualisation. However, the use of Geant4 as a teaching aid in a classroom setting is limited due to the requirement of in-depth knowledge in software. Moreover, even for an experienced user, simulating a simple experiment in the class could be time consuming because one sometimes needs to recompile the code and probably even debug the code before running it. On the other hand, applications like COMSOL\textsuperscript{\textregistered}, which are widely used for multi-physics simulations, offer a user-friendly interface to interactively run a simulation \cite{comsol}. A similar interface for Geant4 would facilitate an experiential learning process for students. 

A tool named Simple has been built with the aim of simplifying particle physics simulations using Geant4. This C++-based application provides a user-friendly interface to create and update geometries, modify beam parameters and also includes a basic data analysis and plotting interface. Moreover, Simple is distributed as a stand-alone application and therefore, compilation of Geant4 libraries is not required.

In this paper, a basic introduction to Geant4 is given, following which the features of Simple are explained along with illustrative examples. While the examples are aimed at teaching students, it must be emphasised here that Simple can also be used for other academic and research purposes.

\section{Geant4: A heuristic point of view}

\begin{figure}[htbp]
    \centering
    \includegraphics[trim={5.5cm 6.5cm 5cm 5cm},clip,scale=0.75]{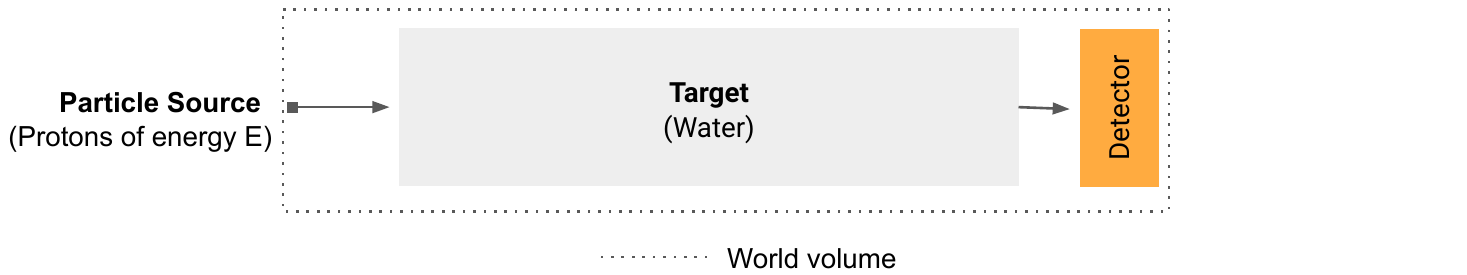}
    \caption{A sample setup to describe the working principles of Geant4.}
    \label{fig:exp}
\end{figure}

The working principle of Geant4 is explained with reference to a simulation of the experimental setup shown in figure \ref{fig:exp}. The ensuing explanation is only approximate and many concepts are ignored or oversimplified; a detailed overview can be obtained from the Geant4 application developer manual \cite{collaboration2012geant4}.

The aim of this experiment is to measure the energy of protons of a certain initial energy $E$ after they pass through a water-tub of specified dimensions (say, $l$, $b$ and $w$). 
\subsection{Shape, material and position definitions}
To simulate this experiment in Geant4, one begins by describing all the objects in the experiment. Here, the objects are a water-tub and a detector. Description of each of these objects requires the creation of the following:
\begin{enumerate}
    \item solid volume (SV)
    \item logical volume (LV)
    \item physical volume (PV)
\end{enumerate}

In the description of the SV, the user defines the shape and dimensions of the object. For the sample simulation, we provide in the SV description for the water-tub, its shape (cube) and dimensions $l$, $b$ and $w$.

Next, in the description of the LV, the material properties along with other physical properties like temperature and pressure of the objects are defined. In our case, the material of our water-tub is water at room temperature. 

In the PV description, we define the positional coordinates of our object. The position coordinates are defined with respect to the centre of the world volume (WV).  The WV is also created the same way the water-tub was created (i.e., creation of SV, LV and PV). The WV can be thought of a laboratory inside which the experimental setup is kept and therefore, the material is usually set to be air. In all Geant4 simulations, the WV is the first volume to be created. The dimensions of the WV are chosen in a way that it just encloses all the objects in the simulation, including the particle source, as indicated by the dashed line in figure \ref{fig:exp}. A bigger WV will result in waste of computational power, slowing down a simulation.

The description of a detector is also done in a similar way. However, certain aspects have to be noted.
In the real experiment, the detector is a complex device with mechanical and electronic subsystems. Nevertheless, in a simulation such details can be ignored unless they substantially influence the measured parameters. Therefore, an object with a simple geometry and a suitable material can be used to serve as a proxy  for the detector. This will also simplify the simulation process to a large extent. In our case, for example, we can choose a cube of air (or vacuum) with a very small thickness to serve as a proxy for our detector. The thickness is made small so that the energy loss in this object is negligible and for other reasons explained in one of the following sections. 

\begin{figure}[htpb]
    \centering
    \includegraphics[trim={2cm 1.5cm 2cm 0cm},clip, scale=0.62]{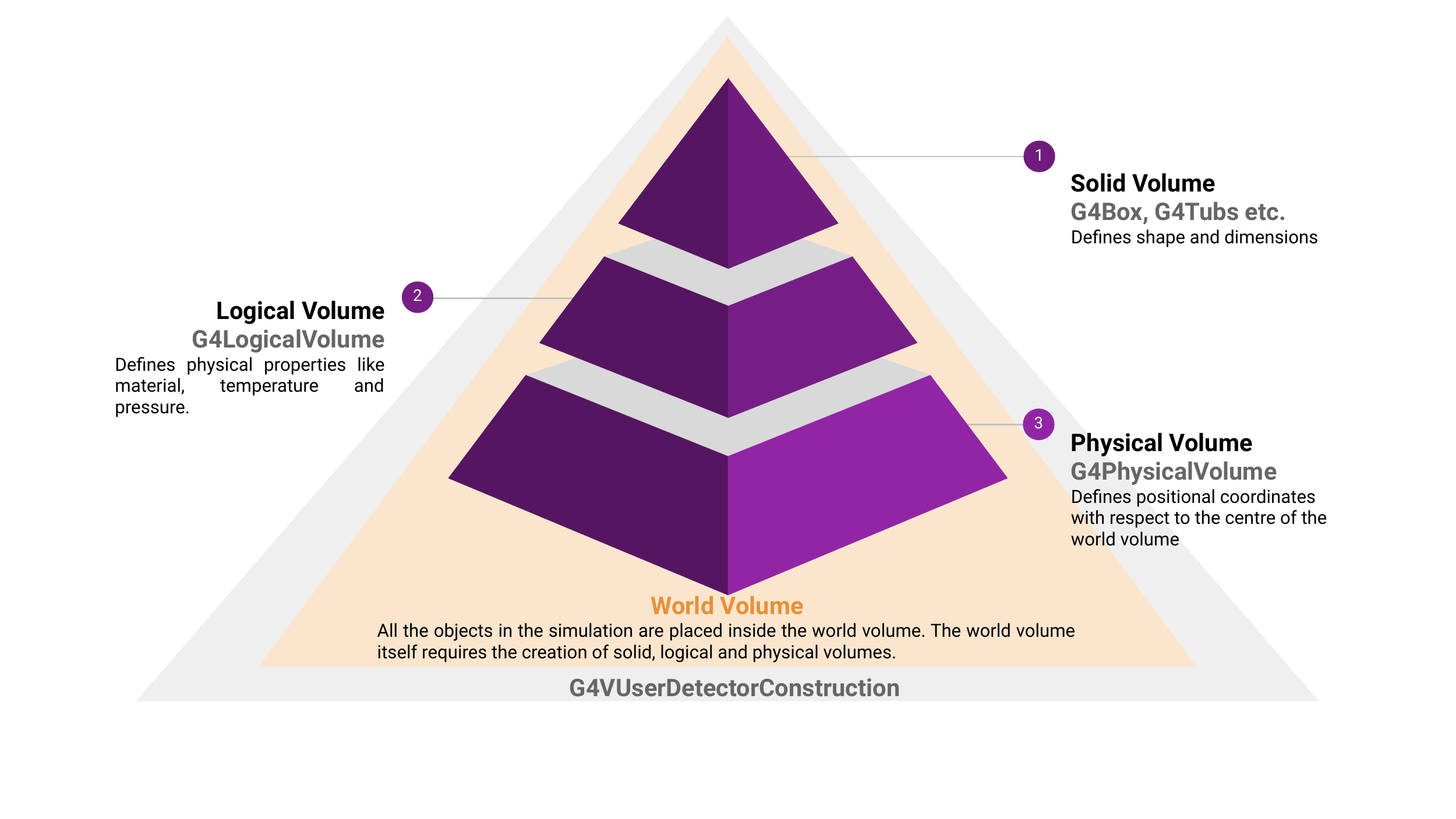}
    \caption{Volumes that need to be created for constructing a geometry in Geant4.}
    \label{fig:g4class}
\end{figure}

The Geant4 class that is used for SV creation depends on the shape of the object. For example, the G4Box class is used for creating a cube while the G4Tubs class is used for a cylinder. The G4LogicalVolume class is used for creating the LV. Geant4 has a comprehensive list of materials that can be further used to create compounds and mixtures, which are used as parameters in constructing the LV. The G4PhysicalVolume class is used for creating the PV. The G4VUserDetectorConstruction class manages the creation of the volumes in a simulation. Figure \ref{fig:g4class} describes the classes required for the creation of the volumes. 

\subsection{Particle source}
The G4VUserPrimaryGeneratorAction class is used for managing the particle source and the G4ParticleGun class is used to define the properties of the particles. Properties like energy, momentum, position and polarization can be set for every particle shot and furthermore, these properties can be randomised or distributed according to specific functions. The G4GeneralParticleSource (GPS) class, in which more features are available to specify the spectral and angular distributions of particles, is another alternative to the G4ParticleGun class. 

In our case, the particle type must be set to proton and the energy will be set to $E$ with a mono-energetic spectral distribution. We also set the momentum vectors such that the particles initially move along the length of the water-tub, which we refer to as the $z$ axis. The angular distribution and the polarisation are not required to be set for this specific case. If the particles are assumed to be originating from a point source,  randomisation of the position (i.e., vertex) is not required. Otherwise, the $x$ and $y$ positions can be generated by pooling numbers from an appropriately weighted random number generator. In addition, the GPS class offers methods to define beam profile specifications, which can be useful in certain applications where the particle source is an accelerator. 

\subsection{Run, event, step and physics lists}
Once the objects and the beam properties are specified, the simulation is started. The simulation begins with shooting a particle from the source. In our case,  the proton with the specified properties moves forward, propagates through the water-tub and exits it. It then enters the detector, exits and propagates until it reaches the world volume. Once the particle exits the world volume, it is no more tracked and the simulation ends here \footnote{It is for this reason that the world volume is usually kept only slightly bigger than the volume that encapsulates all the other volumes. Else, the simulation is carried out in regions that are not relevant.}.  

The preceding description is fairly approximate. It was assumed that the particle travels along a straight path and that it did have sufficient energy to reach the detector, which is not the case always. This sequence of a single particle from its source to the point it exits the world volume is defined as an \emph{event}. A simulation typically consists of many \emph{events}, each one corresponding to a particle with its own properties. For example, the particle position or energy may be different in each \emph{event} so as to simulate a realistic scenario. A group of \emph{events} constitute a \emph{run}. 

The propagation of a particle  in an \emph{event}, through the WV, is made in small units called a \emph{step}. As a particle steps through a volume, one or many physics processes may take place. For example, as a proton steps inside water, it undergoes energy loss in addition to other processes like scattering and/or ionisation. The probability for each of these processes depends on the cross-sections, which are experimentally determined parameters. The cross-sections depend on the properties of the particle like energy, charge, mass, etc., in addition to the material properties of the volume that the particle has stepped in. The cross-section data are separately provided as Geant4 datasets. Essentially, when a particle takes a \emph{step}, the relative probabilities for all the possible processes are computed and one of the possible processes is chosen for the next \emph{step} by using a random number generator. The \emph{step-length} is calculated based on the physics processes, the material properties and the particle properties at the current \emph{step}.  A too long \emph{step-length} would lead to poor accuracy and a very fine \emph{step-length} would increase the computational time.

It is to be also noted that as the particle propagates, it might create other new particles, for example, as a result of ionisation or decay, and the parent particle might disappear. In this case, the newly created particles are also tracked. Hence, an exhaustive list of all the physics processes for all possible particles that are likely to be produced in the simulation is required apriori. This will, however, increase the computation time in a given simulation as cross-section- and/or interpolation-related calculations have to be performed for all these processes. To simplify the calulations, depending on the accuracy expected from the simulations, some processes can be suppressed. For example, a proton is  unlikely to decay and therefore, decay processes can be entirely ignored. Geant4 provides a variety of physics lists, which contain appropriate physics processes for a given experiment, for example, particle decay, underground experiments and hadron therapy.

\begin{figure}[htpb]
    \centering
    \includegraphics[trim={1cm 7.0cm 1cm 0cm},clip, scale=0.55
    ]{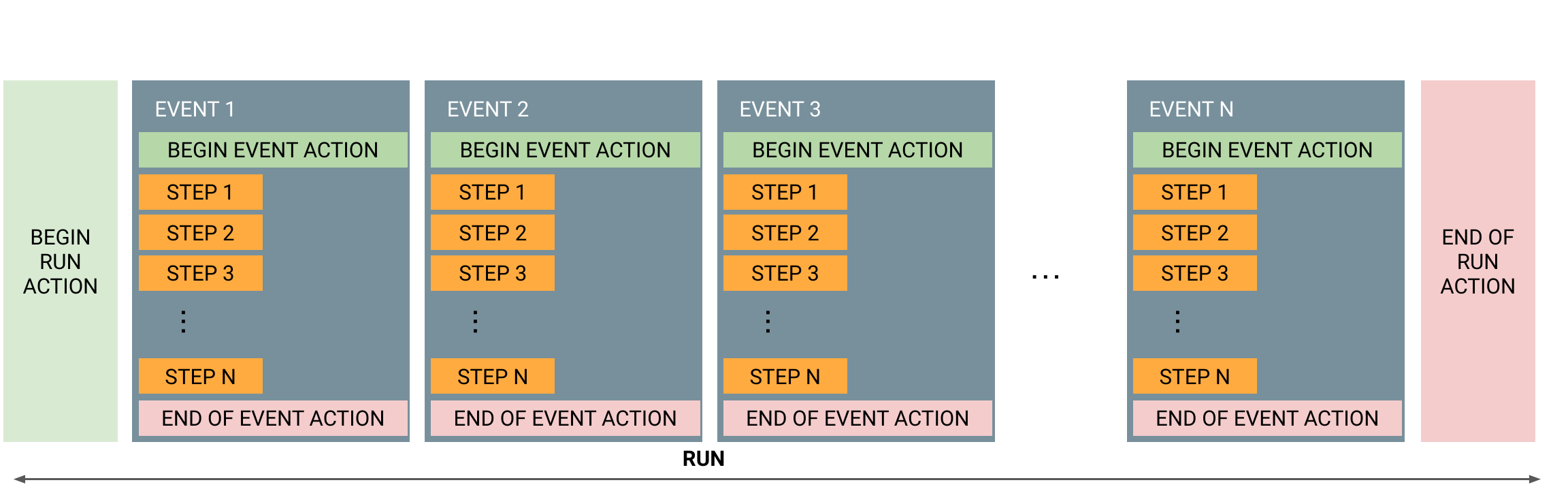}
    \caption{Geant4 concepts of \emph{run}, \emph{event} and \emph{step}. The actions mentioned in the block  are virtual methods that can be modified by users to suit their simulation needs. At the end of each \emph{step}, a method is called, which can be used to obtain the state of a particle at a particular stage of an \emph{event}.}
    \label{fig:runeventclass}
\end{figure}

The Geant4 classes that manage the \emph{run}, \emph{event} and \emph{step} are G4UserRunAction, G4EventAction and G4SteppingAction, respectively. All these classes are initialised in the G4VUserActionInitialization class. The G4UserRunAction class provides methods that are automatically called before and after a \emph{run}, which can be used, for example, to open and close a file. The G4EventAction class also has methods, which are automatically called at the beginning and at the end of an \emph{event}. These methods can be used, for example, to compute quantities on an \emph{event}-by-\emph{event} basis. Figure \ref{fig:runeventclass} shows a block diagram of the concepts of \emph{run}, \emph{event} and \emph{step}.

\begin{figure}[htpb]
    \centering
    \includegraphics[trim={6cm 6.0cm 5cm 5cm},clip,scale=0.85]{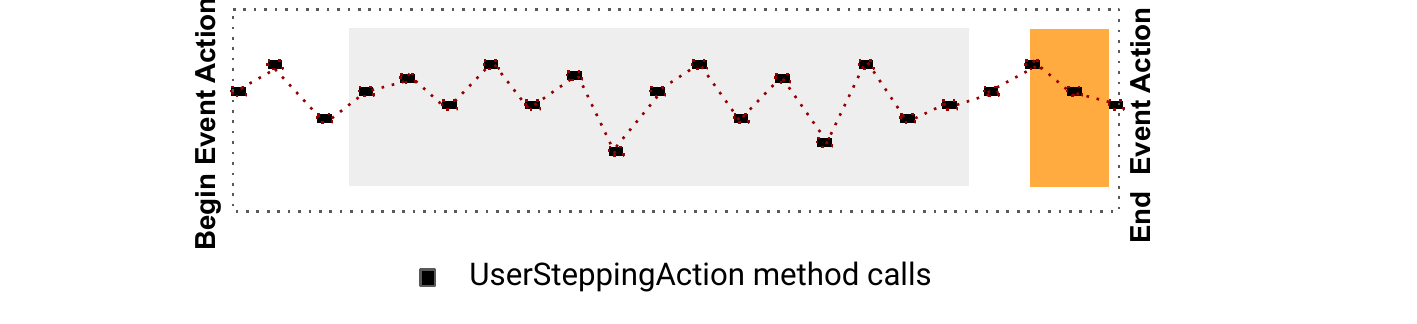}
    \caption{Steps taken by a particle as it travels through the volume. The transition from one point to another point is defined as a \emph{step} in Geant4. At each point, the UserSteppingAction method is called. }
    \label{fig:step}
\end{figure}

Figure \ref{fig:step} shows a typical propagation of a particle through the volume with a dot representing the position after taking a \emph{step}. The distance between two points is the \emph{step-length}. At every point, the UserSteppingAction method (from the  G4UserSteppingAction class) is called, which can be used to intercept the \emph{event} loop and access the particle properties at that point. For our simulation example, in which the energy of a particle after traversing the water volume needs to be determined, we query in each \emph{step}, the volume in which the particle is currently in. In case the particle is in the detector volume, we query the particle energy at that point and store it. As seen from the figure, the particle may take multiple steps inside the detector volume, which may lead to wrong estimation of the energy, especially if the detector material itself leads to substantial energy loss. This can be reduced if the thickness of the detector volume is made as small as the \emph{step-length}.

In many cases, \emph{events} are independent of each other and therefore, the simulation speed can be improved by implementing event-level parallelism using multi-threading. This is possible if Geant4 is compiled with the appropriate settings enabled. 

\subsection{Scoring volume}
Geant4 offers methods to read out data in a format similar to a real experiment. For example, in the case considered so far, a thin slice was placed at the end of the water volume as a proxy for a detector. In a real experiment, this detector could be a pixelated detector enabling pixel-wise data readout. Furthermore, in many cases, the measured quantity (for instance, dose, flux, etc.) is integrated over a specific interval of time. The concept of scoring volume in Geant4 allows one to create such detectors in a simulation. The scoring volume is a mesh placed in the world volume with its shape and pixel binning parameters provided by the user. The scoring volume provides a handle to retrieve physical parameters integrated over a single run with cuts imposed on particle properties provided by the user. Unlike physical volumes, scoring volumes can overlap with other physical volumes in the simulation and do not affect the particle propagation. In addition, Geant4 offers methods to visualise the physical quantity measured by a scoring volume with a colour map. A fine binning of a scoring volume is usually avoided as it substantially increases the computational time.    

\subsection{Macro files}
The simulation parameters can be initialised and controlled through Geant4 specific commands that can be executed on the command line interface (CLI). These can be useful to control the simulation without the need of compiling the code every time a parameter is changed. A set of commands are usually stored in a macro file (.mac extension) and these files can also be called from the CLI. 

The preceding description of Geant4 will be useful in understanding the design of Simple as described in the following section.

\section{Front-end design of Simple}
\begin{figure}[htpb]
    \centering
    \includegraphics[scale=0.18]{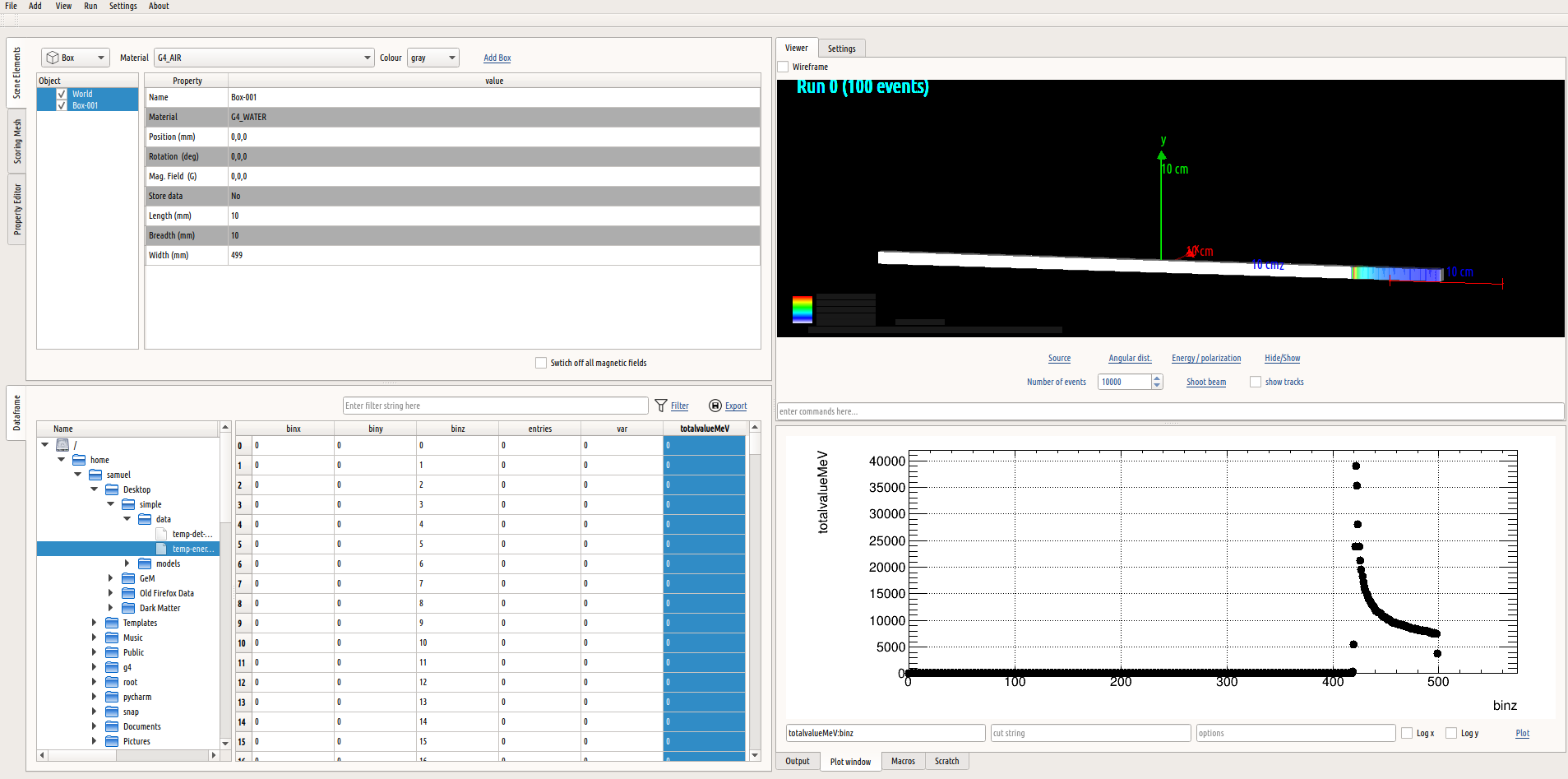}
    \caption{Front-end of Simple. Top-left: interface for creating volumes; Top-right: Geometry and track visualisation; Bottom left: Dataframe viewer; Bottom right: Plots and macros.}
    \label{fig:ui}
\end{figure}

 The user interface (UI) of Simple was designed with the aim of providing fast access to the key elements in simulation. The UI was designed using Qt Creator and the open source version of Qt libraries \cite{qt}. The following section describes the main elements in the UI: 
 \begin{itemize}
    \item \textbf{Geant4 dataset path setting:} This interface is useful for setting the paths of the datasets used in the simulation. The interface also allows one to automatically search and set the paths. Error messages are displayed if the datasets are not found. This setup is done only once on a computer. Figure \ref{fig:dataset} shows a screenshot of this interface.
    \begin{figure}[htpb]
        \centering
        \includegraphics[scale=0.79]{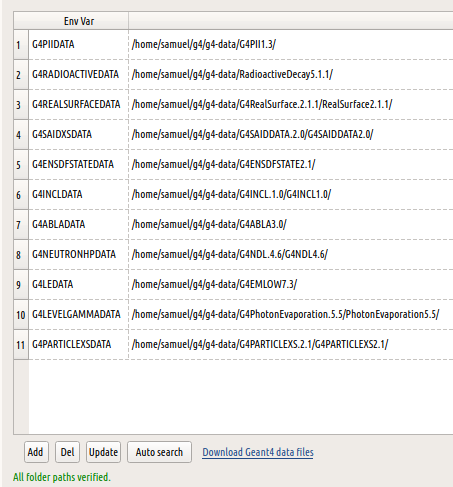}
        \caption{Interface to set and auto-search the Geant4 dataset locations.}
        \label{fig:dataset}
    \end{figure}
     \item \textbf{Physics list setting:} This interface is the first screen to show up. A physics list that is appropriate for the type of simulation to be performed has to be chosen by the user. When a physics list is selected, its description is shown. The physics list cannot be changed during a simulation. Figure \ref{fig:physicslist} shows a screenshot of this interface.
     \begin{figure}[htpb]
         \centering
         \includegraphics[trim={6cm 3.0cm 5cm 0cm},clip,scale=0.75]{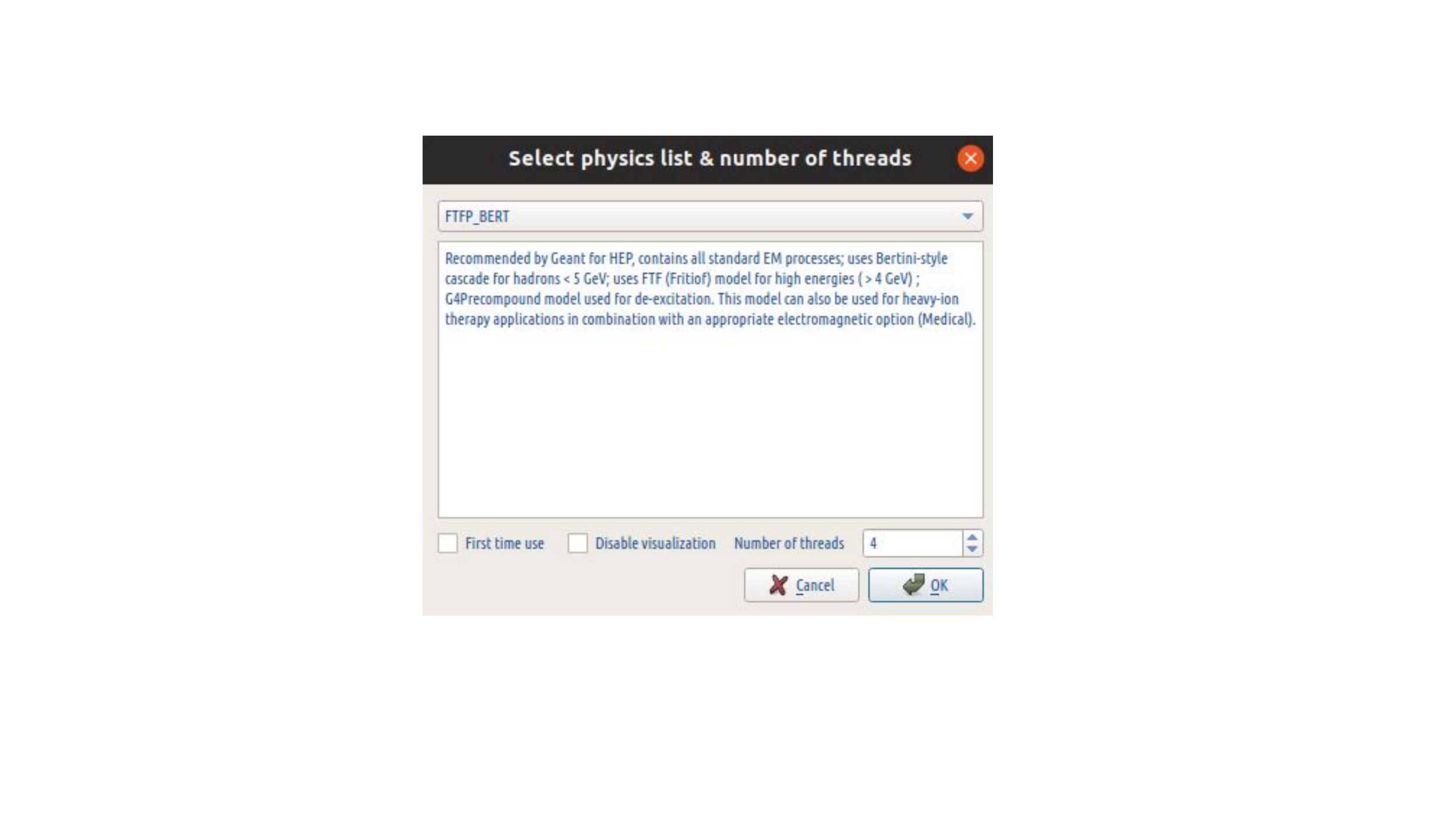}
         \caption{Module for setting the physics list and displaying a descriptive message. The number of threads can also be set in this module.}
         \label{fig:physicslist}
     \end{figure}
    \item \textbf{Volume creation:} This interface is used for creation of solid, logical and physical volumes in one step. A list containing all materials in the Geant4 material database aids in the process of volume creation. At present, only four geometries are available: cube, tube, sphere and wedge. Additionally, an uniform magnetic field can be assigned to the volumes. All parameters of the volumes can be modified when a \emph{run} is not in progress. Each module can be copied and arranged along a specific axis. This is helpful to create an array of detectors, for example. Each volume created can be enabled to readout data of the particles passing through it by enabling the `store data' flag. This feature can be used to make a volume behave like a detector. Figure \ref{fig:volume} shows a screenshot this interface.
    \begin{figure}[htpb]
    \centering
    \includegraphics[trim={0cm 4cm 13cm 2cm},clip,scale=0.75]{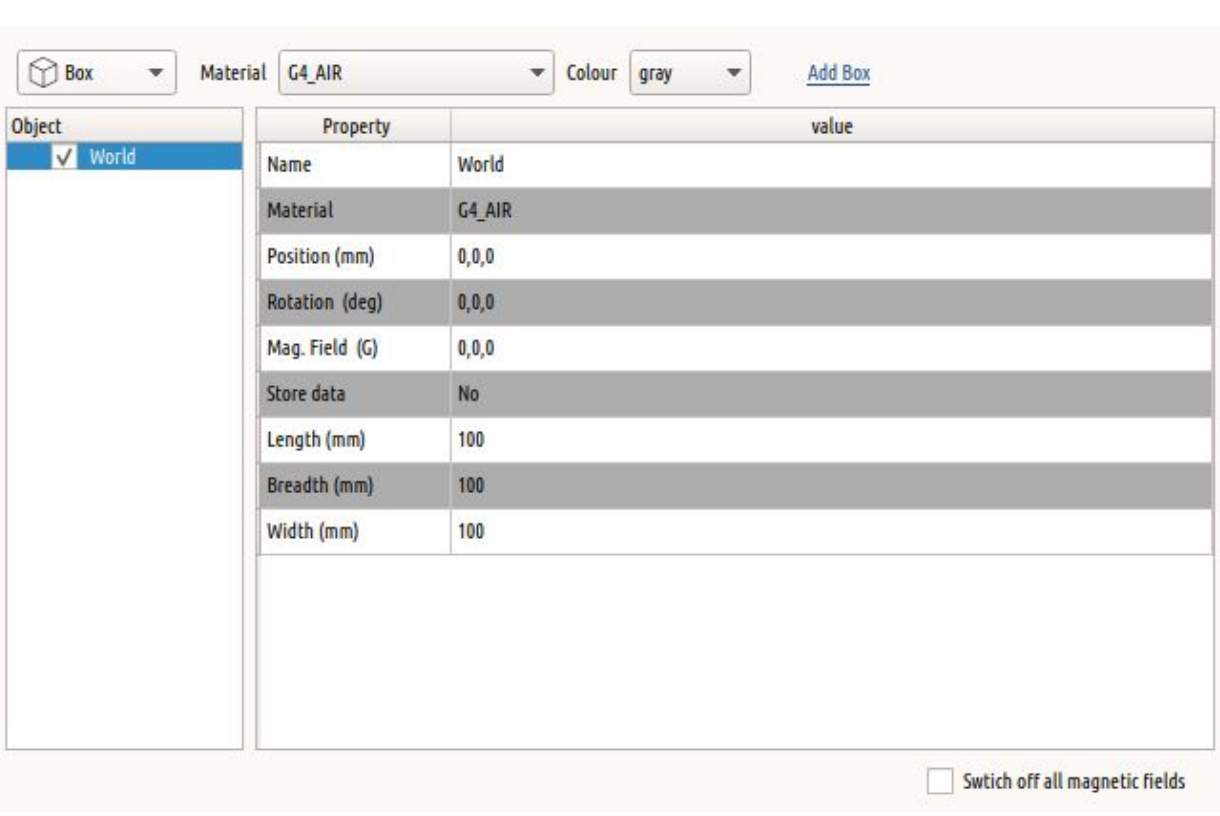}
    \caption{Interface to create and edit volumes in the simulation. A world volume is created by default when the program is started. A menu option is available for every volume from which multiple copies of a volume can be made.}
    \label{fig:volume}
    \end{figure}

     \item \textbf{Scoring volume creation:} This interface helps in creating a box type or a cylindrical type mesh of specified dimension. Each mesh can be assigned a specific physical quantity (chosen from a list) to store. In addition, the data stored in the scoring volumes can  be visualised as a map overlaid on the geometries or separately in a table format in the dataframe viewer. Figure \ref{fig:scoring_volume} shows a screenshot of this interface.
     \begin{figure}[htpb]
    \centering
    \includegraphics[trim={12.5cm 4cm 0cm 2cm},clip,scale=0.75]{volume.pdf}
    \caption{Interface to create scoring volumes. The option to visualise the data from the scoring volumes is available as a context menu.}
    \label{fig:scoring_volume}
\end{figure}

     \item \textbf{Geometry and track visualisation:} This window provides the Qt-based backend provided by Geant4 with provisions to zoom and pan through the world volume. Volumes can also be selected to be hidden or shown. Figure \ref{fig:geometry} shows a screenshot of this interface.
     \begin{figure}[htpb]
         \centering
         \includegraphics[trim={5cm 0cm 5cm 0cm},clip,scale=0.65]{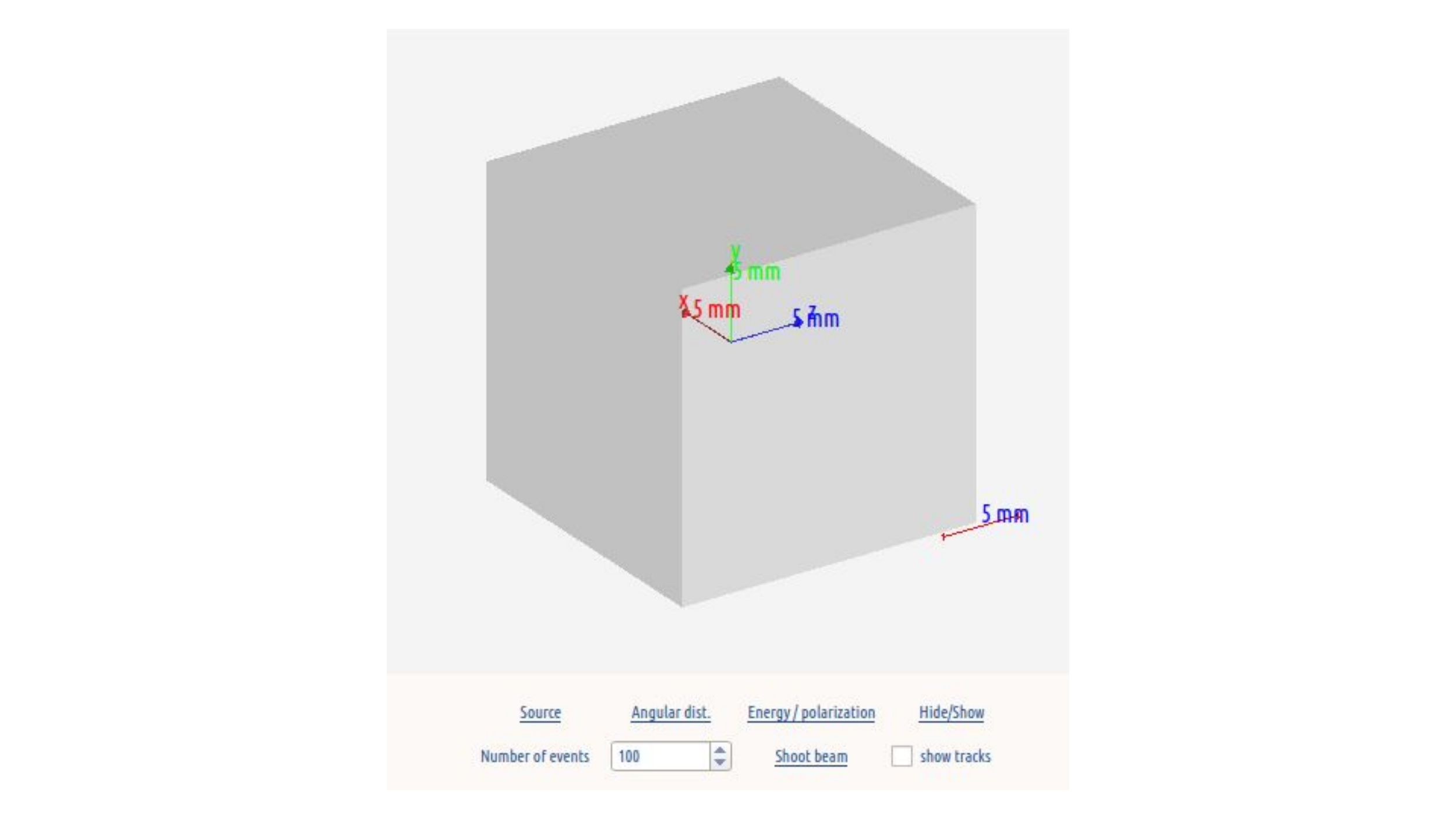}
         \caption{Geometry and track visualisation window. Also shown are the settings available for the particle source parameters.}
         \label{fig:geometry}
     \end{figure}
     
     \item \textbf{Particle source setting:}  This interface provides access to particle parameters including angular and spectral distribution and polarisation. Figure \ref{fig:geometry} (bottom part) shows a screenshot of this interface. 
      \item \textbf{Data analysis:} A dataframe viewer shows the output from the simulation, both from the scoring volumes and volumes in which data storing is enabled, in a table format. The data in the table can be plotted on a ROOT-based canvas. Figure \ref{fig:datanalysis} shows a screenshot of this interface.
      \begin{figure}[htpb]
          \centering
          \includegraphics[trim={2cm 3cm 0cm 2cm},clip,scale=0.60]{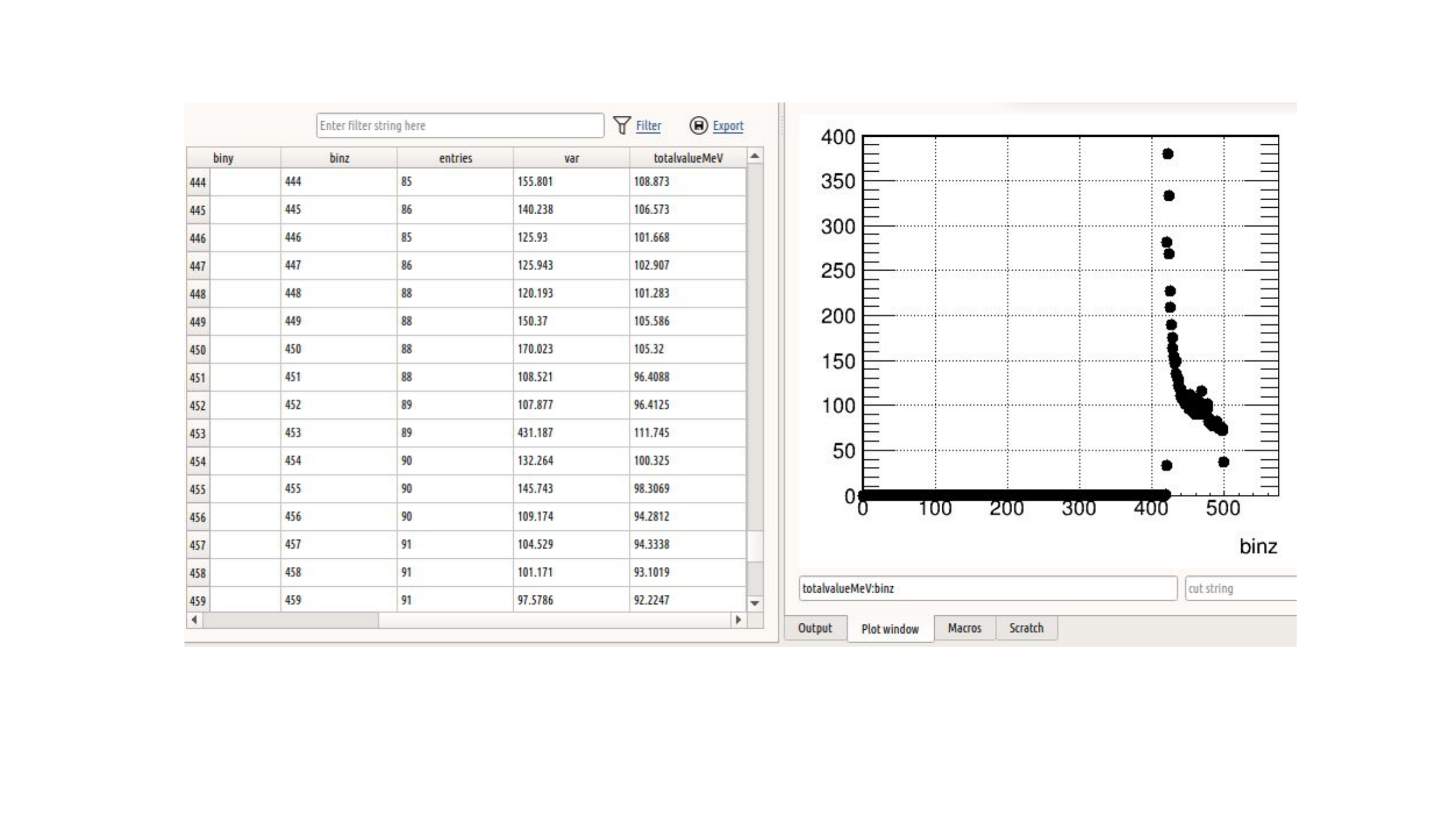}
          \caption{Dataframe viewer and the plotting canvas in Simple.}
          \label{fig:datanalysis}
      \end{figure}
     \item \textbf{New material creation:} This module helps in creating new materials and compounds using existing materials in the Geant4 database. The new material is saved to the database so that it is available for future use. In addition, material properties can also be specified in this module. Figure \ref{fig:material} shows a screenshot of this interface.
     \begin{figure}[htpb]
         \centering
         \includegraphics[trim={1cm 0cm 1cm 0cm},clip,scale=0.50]{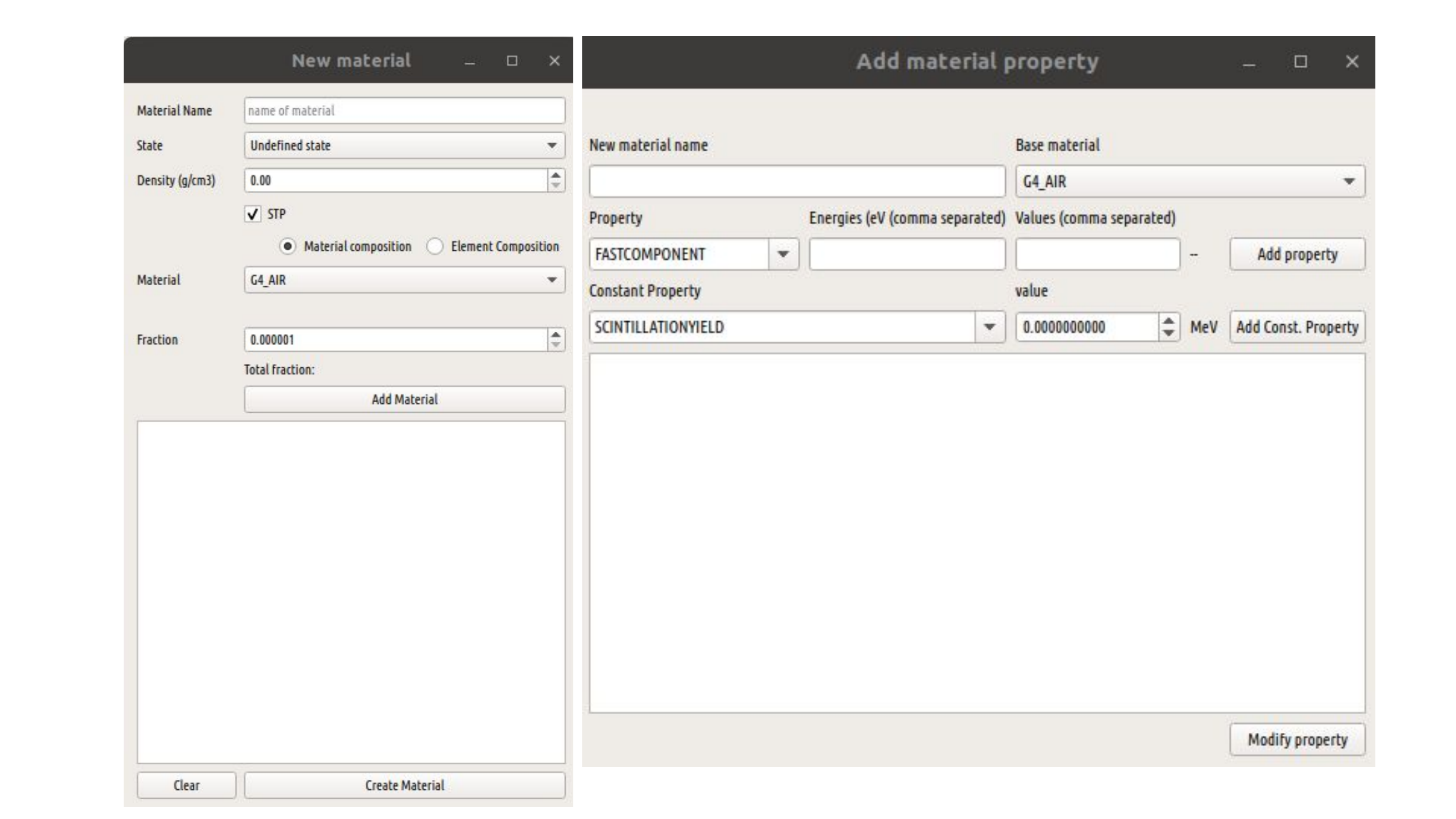}
         \caption{Interface for material and property editing in Simple.}
         \label{fig:material}
     \end{figure}
     \item \textbf{Geant4 macro execution:} A separate window in which a macro can be written and executed, helps in managing and automating a simulation process.
     \item \textbf{Project management:} All volumes, particle source settings and macros can be saved for future use. 
\end{itemize}
 A screenshot of the application with all the elements described above is shown in figure \ref{fig:ui}.

\section{Workflow of Simple}

The simulation workflow begins by selection of an appropriate physics list and the number of threads to be used in the simulation. Next, the required volumes are created in the appropriate interface.  In case a material is not found in the database, it is be created and inserted in the database before proceeding to create a volume.  

A world volume with default parameters is created when the program is initialised and other volumes are sequentially inserted inside the world volume. The material, positional and dimensional aspects can be edited on a single interface once a volume has been inserted. The same interface allows the user to assign a magnetic field to a volume. The interface has the provision to enable data storing for every volume. As mentioned earlier, if this feature is enabled, when a particle crosses the corresponding volume, its properties are stored in an output file. This will be useful to track the properties of the particle at a given point or to convert a volume into a detector. The user can also select the properties to be written in the output file.  As an alternative, the user can also create a scoring volume and select the parameters to be integrated and written in an output file. 

Next, the particle source properties like position, angular and spectral information are set, following which the number of \emph{events} is entered and a \emph{run} is started. The simulation will run in the background for  a
time duration that depends  on the complexity of the simulation and the number of particles shot. Once the simulation ends, the particle tracks are displayed on the screen, if enabled. The data for the volumes for which data storage was enabled, will be saved in ROOT format in a directory named `data' in the same folder as that of the application. If a scoring volume was created, the data from these volumes are also saved. These files can be selected on the file selection interface and the contents are displayed in a table in the dataframe viewer. The data displayed in the dataframe can also be plotted on a ROOT canvas.

\begin{figure}[htpb]
    \centering
    \includegraphics[trim={1cm 3cm 2cm 2cm},clip,scale=0.65]{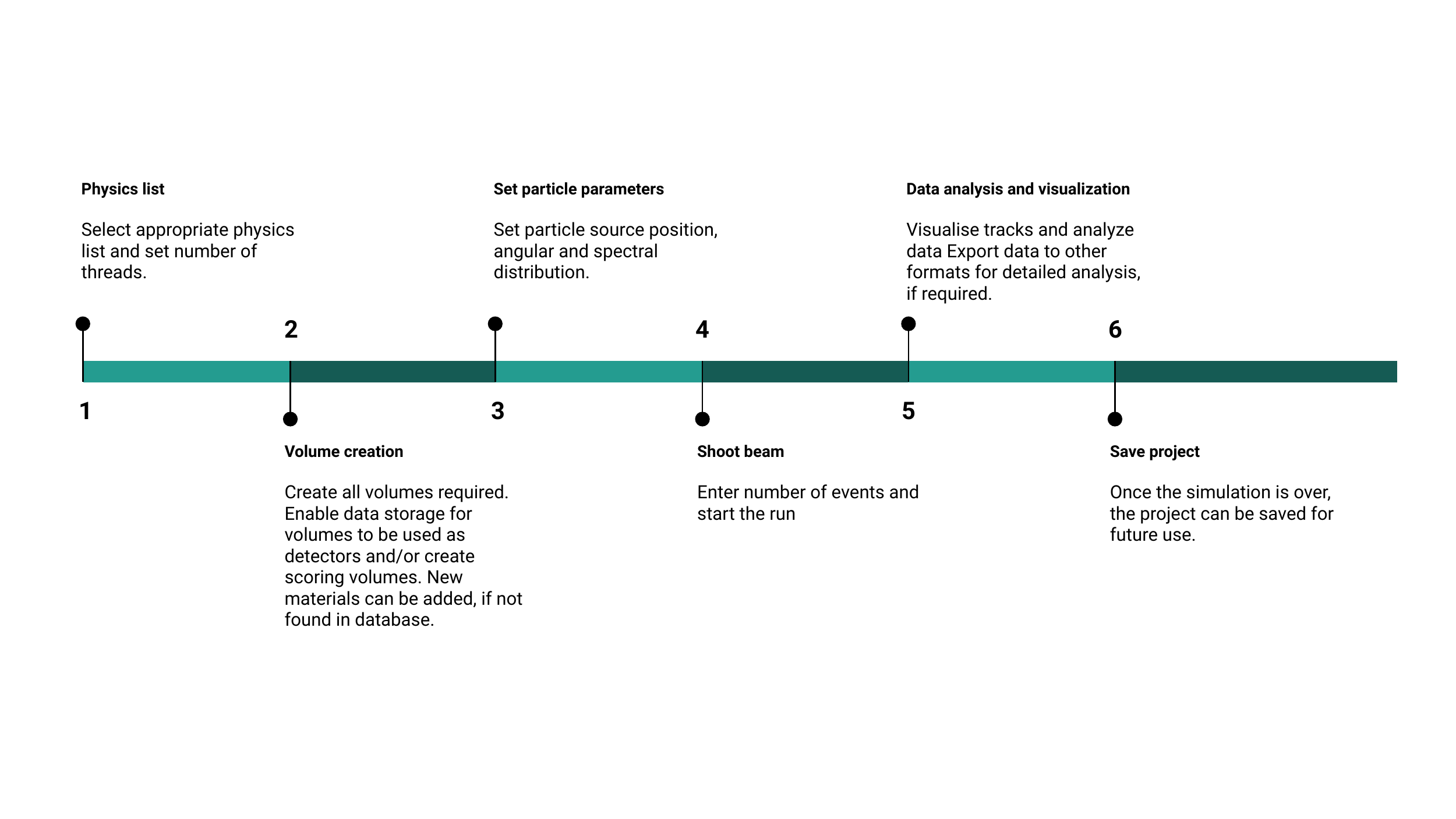}
    \caption{Typical simulation workflow in Simple.}
    \label{fig:workflow}
\end{figure}

The simulation can be repeated after adjusting the parameters of the volumes or the particle source. The workflow is illustrated in figure \ref{fig:workflow}. The entire project can be saved in a custom file format for future use.

\section{Sample applications}
\subsection{Decay spectrum of  $^{60}_{27}Co$}
\begin{figure}[htpb]
    \centering
    \includegraphics[trim={3cm 5cm 3cm 3cm},clip,scale=0.75]{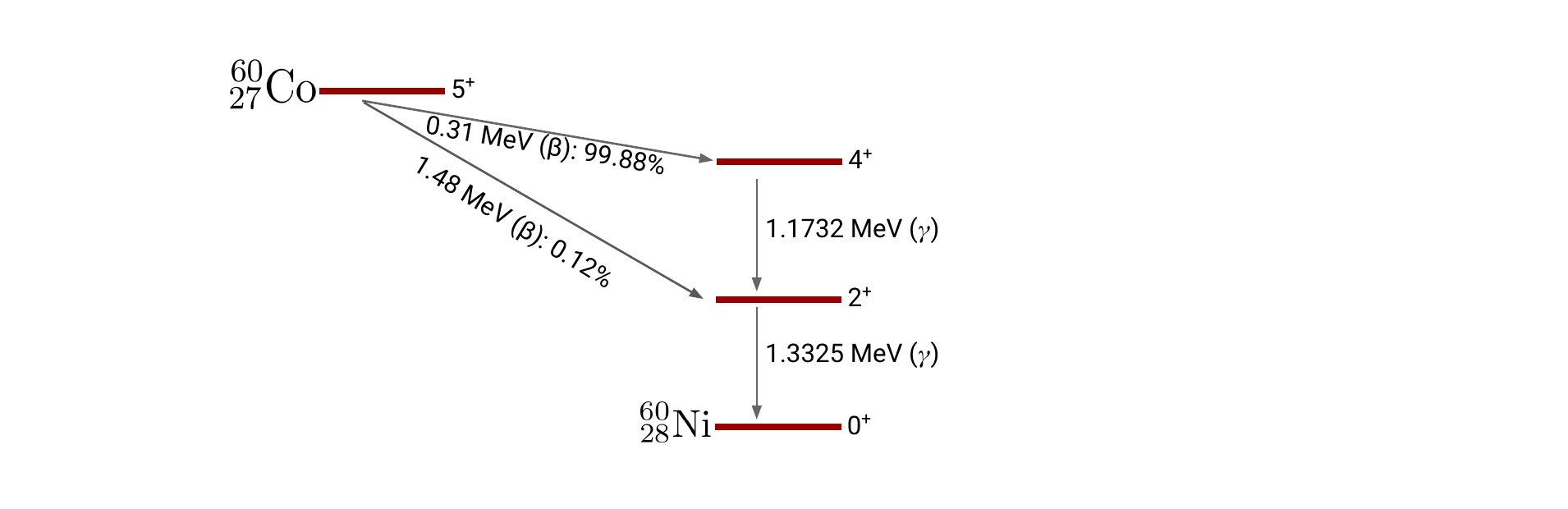}
    \caption{Decay scheme of $^{60}_{27}Co$.}
    \label{fig:cobalt}
\end{figure}

We start with the simulation of $^{60}_{27}Co$ decay, where the goal is to study the various decay modes of $^{60}_{27}Co$, shown in figure \ref{fig:cobalt}. There are two channels by which $^{60}_{27}Co$ decays to the ground state of $^{60}_{28}Ni$. In the first channel, a $\beta$ particle of  energy less than 0.31 MeV is emitted, followed by the emission of 1.1732 MeV $\gamma$, which is further followed by 1.3325 MeV $\gamma$ emission. 
In the second channel,  a $\beta$ particle of energy less than 1.48 MeV is emitted, followed by the emission of 1.3325 MeV $\gamma$. The $\beta$ particles are emitted along with their associated electron anti-neutrinos such that the total energy of the $\beta$ particle and that of the neutrino is 0.31 MeV and 1.48 MeV for the first and second channels, respectively. The half-life of $^{60}_{27}Co$ is 5.27 years ($\lambda = 0.1315$ $yr^{-1}$) and the branching ratio for the decay through the first channel is 99.88\% \cite{cobalt}.

\subsubsection{Simulation setup}
The first step in the simulation is the choice of an appropriate physics list. The `Radioactive decay' option is best suited for this simulation. Next, a world volume of appropriate size is created, at the centre of which a hollow sphere with negligible shell thickness is placed. The sphere is meant to act as a detector, and hence the `store data' flag must be enabled for this volume. For this example, the particle source will be $^{60}_{27}Co$ and the appropriate atomic number and the mass numbers are selected from the particle source interface. The spectral distribution is not required as the kinematics is already governed by physical principles. The angular properties are also not required in this case.  The particle source is also kept at the centre of the world volume such that the sphere surrounds it. This setup will enable particles emitted in all directions to be readout by the detector. 

\begin{figure}[htpb]
    \centering
    \includegraphics[trim={3cm 5cm 3cm 0cm},clip,scale=0.70]{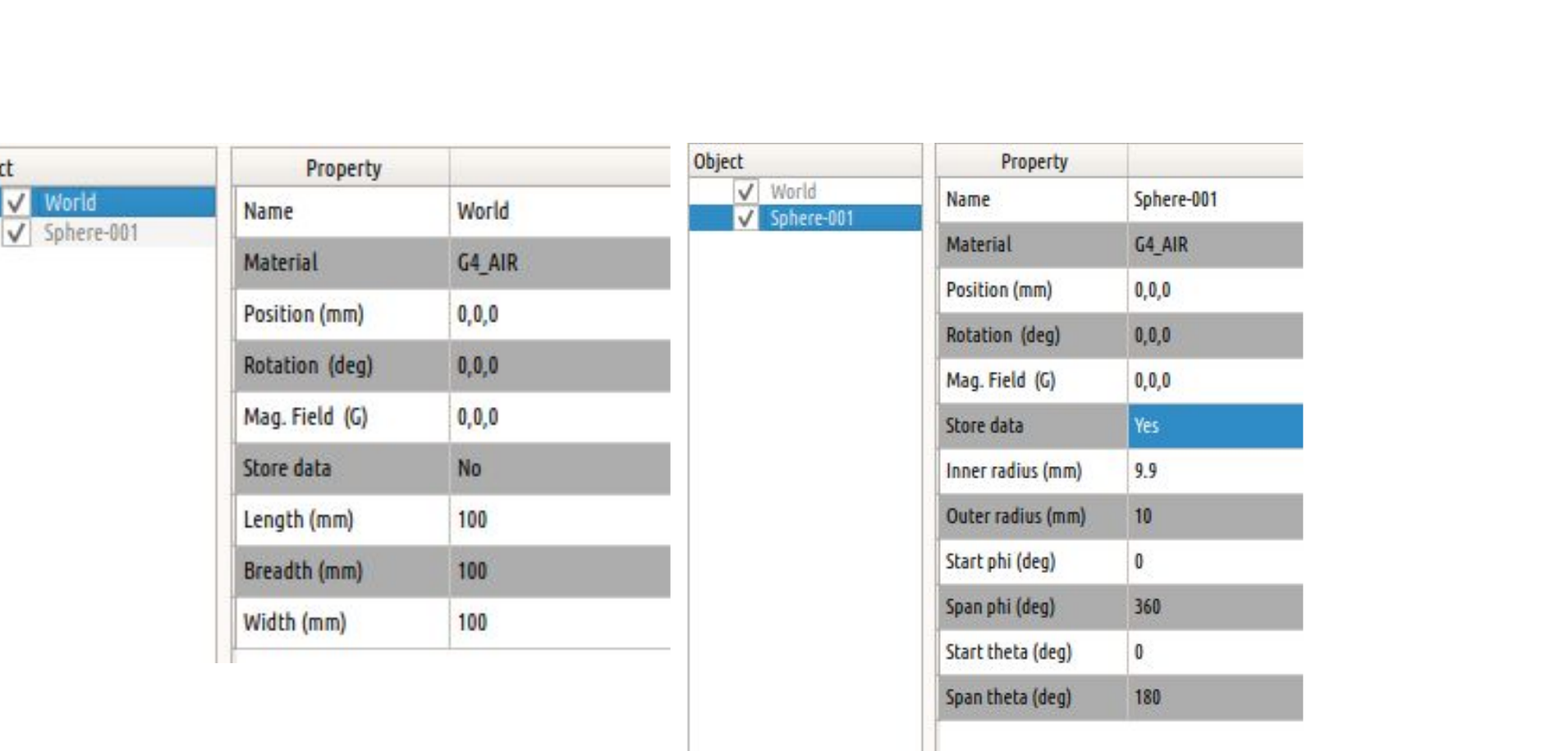}
    \caption{Parameters used for creation of the world volume and sphere for $^{60}_{27}Co$ decay spectrum simulation. The `store data' flag is enabled for the sphere and therefore, whenever a particle crosses it, the particle properties are stored.}
    \label{fig:co-volume}
\end{figure}

\begin{figure}[htpb]
    \centering
    \includegraphics[trim={5cm 0cm 3cm 0cm},clip,scale=0.78]{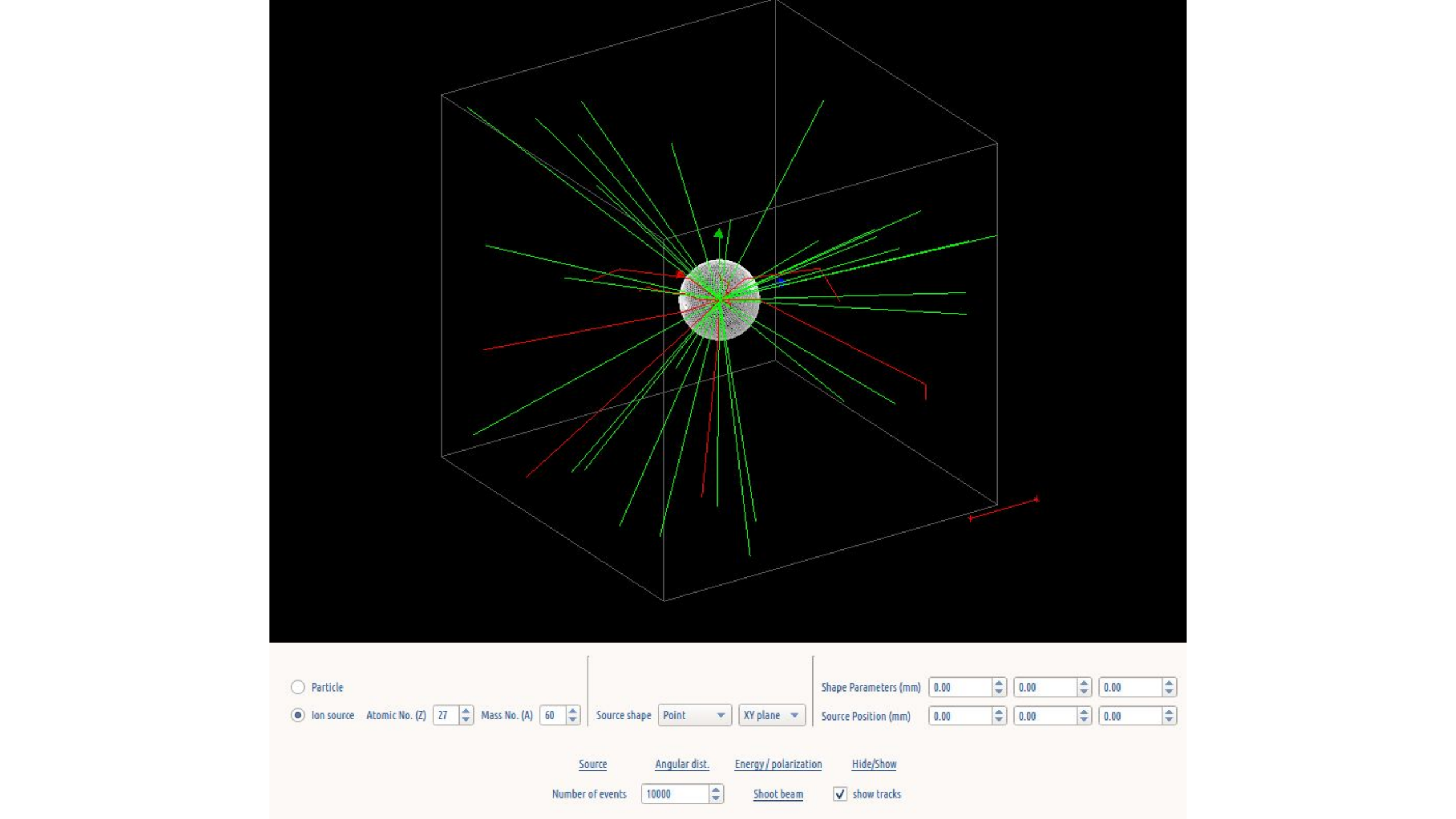}
    \caption{Screenshot of Simple with particle source parameters and sample tracks. For radioactive sources, the atomic number and mass number have to be provided. The colour of the tracks represents the charge of the particles. The Geant4 default colours are red, green, blue for negative, neutral and positive charge, respectively.}
    \label{fig:co-particles}
\end{figure}

The parameters for the world volume and the sphere are shown in figure \ref{fig:co-volume}. A screenshot of Simple showing the particle source parameters and a visualisation of few events is shown in figure \ref{fig:co-particles}.

\subsubsection{Data analysis}
The simulation is run with about 10,000 events. The resulting output file is automatically opened at the end of the \emph{run} and can be viewed in the dataframe viewer. The data typically consist of the columns shown in table \ref{tab:my-table}.

\begin{table}[htpb]
\begin{tabular}{|l|l|}
\hline
Short name & Description                           \\ \hline
eid        & An unique number given to an \emph{event} \\ \hline
pid        & PDG-based particle code            \\ \hline
detectorId & A string to identify a detector    \\ \hline
posx, posy, posz & Coordinates of the particle at the time of crossing the detector         \\ \hline
globTime   & Time since the current \emph{event} began \\ \hline
propTime   & Time since the current track began \\ \hline
localTime        & Time in the rest frame of the particle since the track began             \\ \hline
px, py, pz         & Three momenta of the particle at the instant of crossing the detector    \\ \hline
tot\_E, tot\_KE  & The total energy and kinetic energy at the time of crossing the detector \\ \hline
\end{tabular}
\caption{Parameters stored in the output file in a typical simulation in Simple. The user can disable or enable in the settings the storage of specific parameters. The short name is used for plotting the corresponding parameters.}
\label{tab:my-table}
\end{table}

\begin{figure}[htpb]
    \centering
    \includegraphics[trim={2cm 0cm 3cm 0cm},clip,scale=0.70]{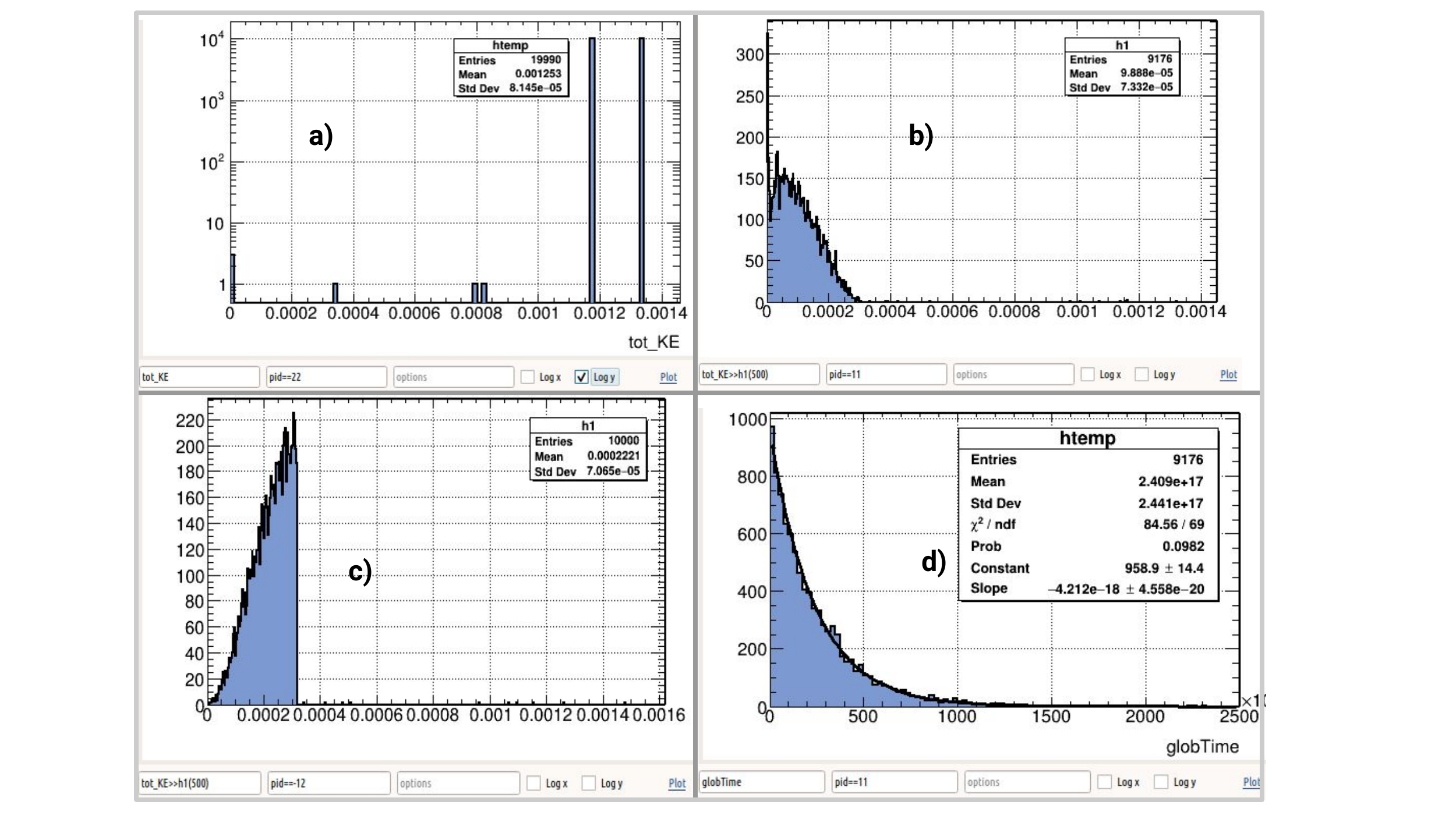}
    \caption{Plots generated using the simulated data of $^{60}_{27}Co$. a) Energy spectrum of the photons (pid=22). b) Energy spectrum of electrons. c) Energy spectrum of electron anti-neutrinos (pid=-12). d) Time distribution of electrons (pid=11). The default units for time and energy are ns and GeV, respectively. The PDG particle ids for photon, electron and anti-neutrino are 22, 11 and -12, respectively.}
    \label{fig:co-plots}
\end{figure}

With the aid of the in-built plotting tool, few analyses can be directly performed. To estimate the branching ratio from the simulation, one can plot the energy distribution of photons (pid 22). From the photons in the first and second channel, we can expect two peaks at 1.1732 MeV and 1.3325 MeV, respectively. This distribution, as shown in the plotting tool in Simple, is presented in figure \ref{fig:co-plots} a. The energy distribution of photons is plotted by entering the string \emph{tot\_KE} in the plot string field and \emph{pid==22} in the cut string field. The plot also shows that the total entries is 19,990, from which the branching ratio can be estimated as follows.

The total number of events in the simulation is 10,000. If $N_1$ and $N_2$ denote the number of events in the first and second channels, respectively, $N_1+N_2 =10000$. However, since in the first channel two photons are emitted whereas in the second channel only one photon is emitted, we can write $2N_1+N_2 =19990$. From these two equations, we determine that the branching ratio for the first channel is 99.90\%, which is close to the true value. It is important to note that the $y$ axis of the plot (Fig. \ref{fig:co-plots} a) is shown in logarithmic scale to emphasise the bins with low counts. Though only mono-energetic photons at the two energies mentioned are expected (as highlighted by the two strong peaks), these photons might sometimes undergo other processes like compton scattering, ejecting an additional electron in the process. Depending on the scattering angle, the electron can take up most of the energy, leaving the photon with less energy. The low count bins are due to such effects. 

Next, the energy spectrum of the electron and electron anti-neutrino can be plotted by using the string \emph{tot\_KE} in the plot string field and \emph{pid==11} in the cut string field for electrons and \emph{pid==-12} for anti-neutrinos. The corresponding distributions are shown in figures \ref{fig:co-plots} b and \ref{fig:co-plots} c, respectively. From the distribution for electrons, we observe that the spectrum settles to zero at about 0.3 MeV, which is the true end-point energy of $\beta$ in the first channel. The second channel spectrum will not be visible as the branching ratio is very small. Next to note is that the number of entries is 9,176 for electrons and 10,000 for anti-neutrinos. This is due to the fact that the electrons, which do not have sufficient energy, will not reach the detector and will be lost while the neutrinos being almost massless cannot stay at rest and travel at the velocity of light. 

The half-life can be determined as follows. The plot string is set as \emph{globTime} and the cut string is set to \emph{pid==11}. The spectrum is then fitted to an exponential function using the fit panel (ROOT-based) that can be triggered from the context menu on the plotting canvas. The fit parameters as estimated from the fit are displayed on the canvas, as shown in figure \ref{fig:co-plots} d. The slope parameter corresponds to the decay constant $\lambda$ from which the half-life can be estimated as $0.63/\lambda = 1.64\times10^8$ s $= 5.2$ a. 

\subsection{$\alpha$ scattering off a gold foil}
One of the important discoveries in nuclear physics was the presence of the nucleus, based on the famous Rutherford scattering experiment. The simulation of this experiment is straightforward and the analysis gives a first-hand experience on calculating important quantities like cross-sections. The experimental details can be found in many literature but for the ensuing discussion, we will follow the material in \cite{wangrutherford}. 

A source of $\alpha$ particles is kept in front of a thin gold foil and a particle counter is placed at various angles to measure the scattered particle count. In the experiments initially performed by Geiger and Marsden, it was observed that $\alpha$ particles scattered through large angles. That a massive but small positively charged entity (now called the nucleus) is the reason for such large scatterings, was the proposal put forth by Rutherford. The following expression of the cross-section for such processes was also derived by him:
\begin{equation}
    \frac{d\sigma}{d\Omega} = \left(\frac{ZZ'\alpha\hbar c}{4E}\right)^2\frac{1}{sin^4(\theta/2)}
\label{eqn:rcs1}
\end{equation}
where, $Z$ is the charge of the particles, $Z'$ is the charge of the target nuclei, $E$ is the kinetic energy of the particle and $\theta$ is the scattering angle. $\alpha$ is the fine structure constant, i.e., 1/137.

Also, by definition:
\begin{equation}
    \frac{d\sigma}{d\Omega} = \left(\frac{I_\theta}{I_o}\right)\frac{A}{d\Omega \cdot N_A \cdot \rho \cdot t}
    \label{eqn:rcs2}
\end{equation}
where, $I_\theta$ is the intensity (counts) of $\alpha$ particles at an angle $\theta$ with respect to the incident direction, $I_o$ is the integral number of particles from the source, $A$ is the atomic weight of the target nucleus, $N_A$ is Avogadro’s number, $d\Omega$ is the solid angle of the detector and $\rho$ is the density of the target foil and $t$ is its thickness. 

By equating \ref{eqn:rcs1} and \ref{eqn:rcs2}, we get:
\begin{equation}
\left(\frac{I_\theta}{I_o}\right)=\frac{d\Omega \cdot N_A \cdot \rho \cdot t}{A}  \left(\frac{ZZ'\alpha\hbar c}{4E}\right)^2\frac{1}{sin^4(\theta/2)}
    \label{eqn:rcs3}
\end{equation}
The LHS is a measurable quantity and, as seen above, is proportional to $cosec^4(\theta/2)$. In the RHS, the proportionality constant can be  theoretically estimated or by fitting $cosec^4(\theta/2)$ to experimentally measured data of $I_\theta/I_o$ as a function of $\theta$.  
With the aid of the simulation, it will be possible to show the important results of the actual experiment. 
\subsubsection{Simulation setup}
The FTFP-BERT physics list is used for this simulation. A gold foil of thickness 0.004 mm and length and width 10 mm each is kept inside a hollow sphere of outer radius 10 mm and inner radius 9.9 mm. This sphere is considered as a detector in the simulation, from which we can count the number of particles impinging on it. The position of the gold foil is offset by 2 mm from the origin along the $z$ axis.
This setup is placed inside a world volume, which is a cube of dimension 100 mm $\times$ 100 mm $\times$ 100 mm. The material for the world volume and sphere is set to vacuum (G4\_Galactic material in Geant4) since any other material would cause the $\alpha$ particles to scatter and attenuate. An $\alpha$ source, of energy 5 MeV, is kept at the centre of the world volume and the beam direction is set along the $z$ axis such that all particles move in the same direction to hit the face of the gold foil. 
The particles scatter off the gold foil and reach the sphere where their positions are recorded. This positional information is used to calculate the scattering angle from which $I_\theta$ can be estimated.\footnote{
The experimental setup is different from this simulation setup. In particular, a detector with a limited solid angle and which is rotatable about an axis was used in the original design. The detector is rotated by various angles to measure the counts at those angles, which yields $I_\theta$.}

\begin{figure}[htpb]
    \centering
    \includegraphics[trim={5cm 2cm 3cm 2cm},clip,scale=0.65]{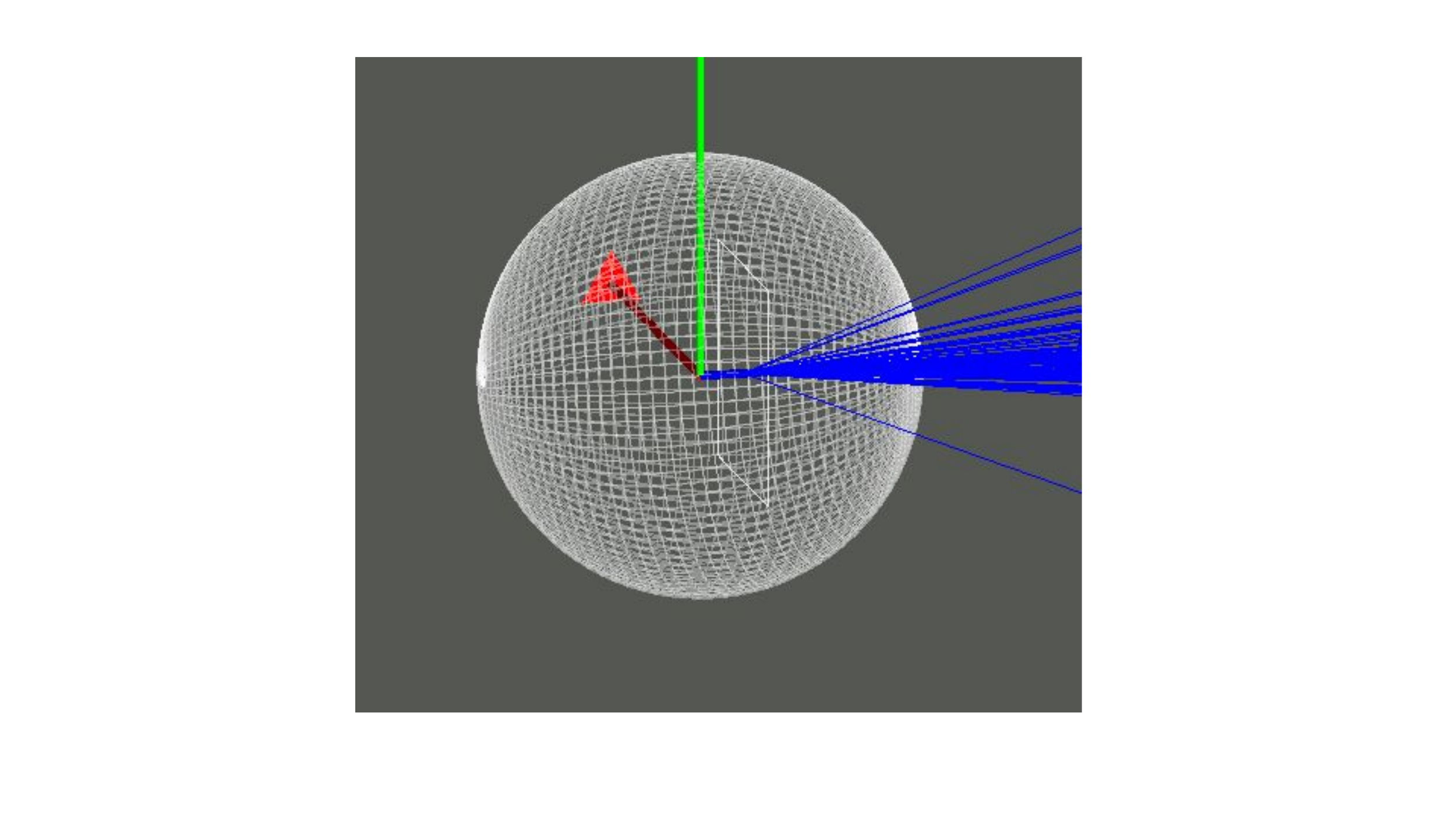}
    \caption{Geometry of the $\alpha$ particle scattering experiment. The particles start from the origin, hit the gold foil (shown as a rectangle) and get scattered to reach the sphere, where the position information is recorded. }
    \label{fig:alphascattering}
\end{figure}

The parameters of the volumes used in the simulation are summarised in table \ref{table:alphaparameters} and figure \ref{fig:alphascattering} shows the tracks produced by the $\alpha$ particles.
\begin{table}[]
\begin{tabular}{lll}
Volume              & Dimensions                             & Position (mm) \\ \hline
World volume (cube) & 100 mm $\times$ 100 mm $\times$ 100 mm & 0, 0, 0  \\
Sphere              & Radius, outer: 10 mm, inner: 9 mm  & 0, 0, 0  \\
Gold foil (cube)    & $l=b=10$ mm; w (along $z$): 0.004 mm   & 0, 0, 2  \\ \hline
                       
\end{tabular}
\caption{Parameters used for the simulation of $\alpha$ particle scattering. The world volume and sphere materials are set to vacuum (G4\_Galactic material in Geant4).}
\label{table:alphaparameters}
\end{table}

\subsubsection{Data analysis}
\begin{figure}[htpb]
    \centering
    \includegraphics[trim={2cm 2cm 0.8cm 2cm},clip,scale=0.55]{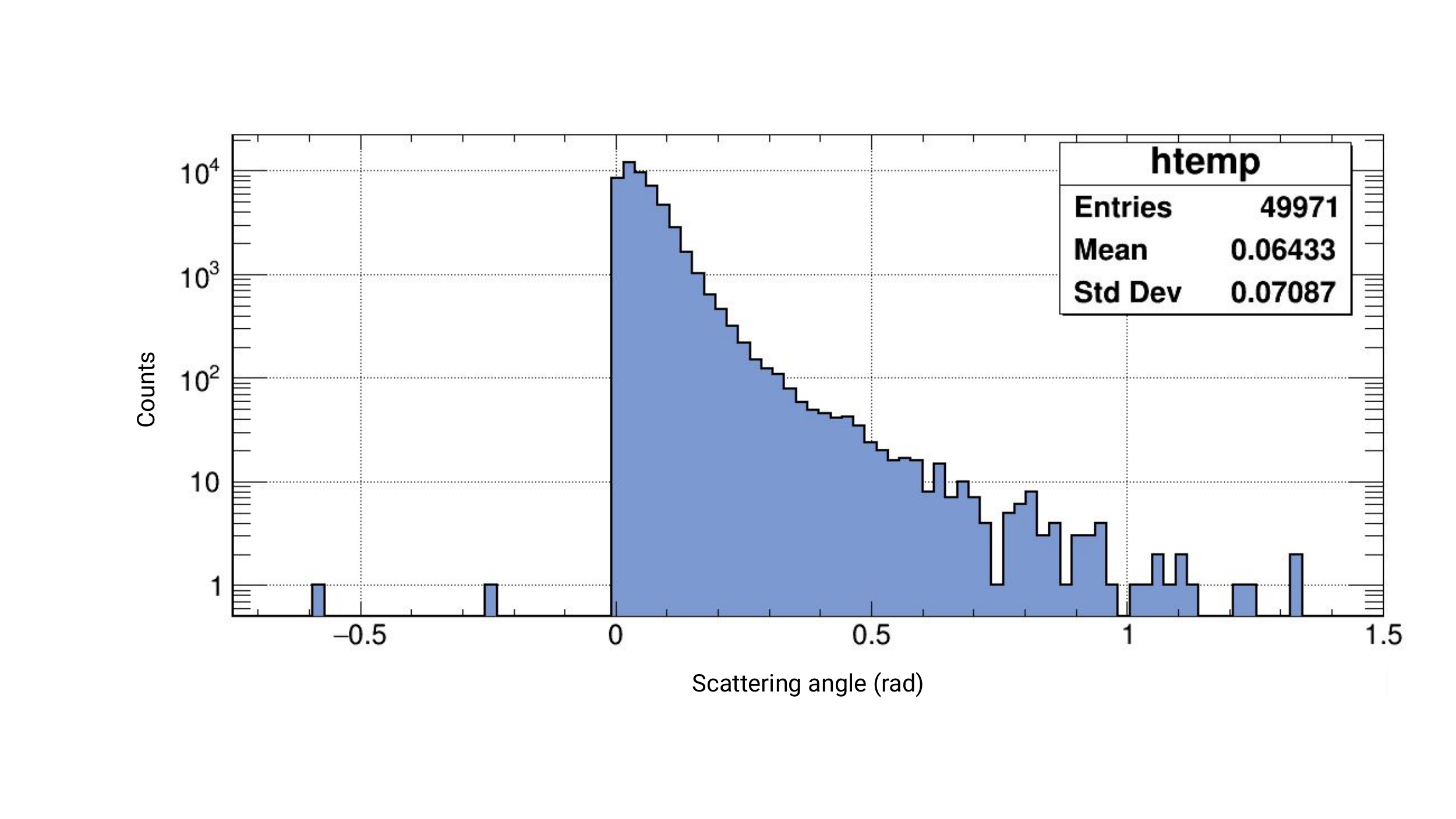}
    \caption{Distribution of the scattering angles in the $\alpha$ particle scattering simulation obtained using equation \ref{eqn:theta}, replacing the terms x, y, z by their equivalent strings `posx', `posy', `posz' in the plotting tool. The two bins in the negative side are due to backscattering.}
    \label{fig:alphascatteringplot}
\end{figure}

Since the sphere in the simulation is considered as a detector, the information of particles crossing it is stored in a file, which is automatically opened at the end of a \emph{run}. For the demonstration \emph{run} that is discussed here, about 50,000 events were generated. The scattering angle was calculated using the following formula:
\begin{equation}
    \theta = \left(\frac{abs(z)}{z}\right) cos^{-1}\left(\frac{\sqrt{x^2+y^2}}{\sqrt{x^2+y^2+z^2}}\right)
    \label{eqn:theta}
\end{equation}
where, the first factor is used to determine the sign of the scattering angle. The histogram of the scattering angle is then obtained, as shown in figure \ref{fig:alphascatteringplot}.

\begin{figure}[htpb]
    \centering
    \includegraphics[trim={2cm 2cm 0.8cm 2cm},clip,scale=0.55]{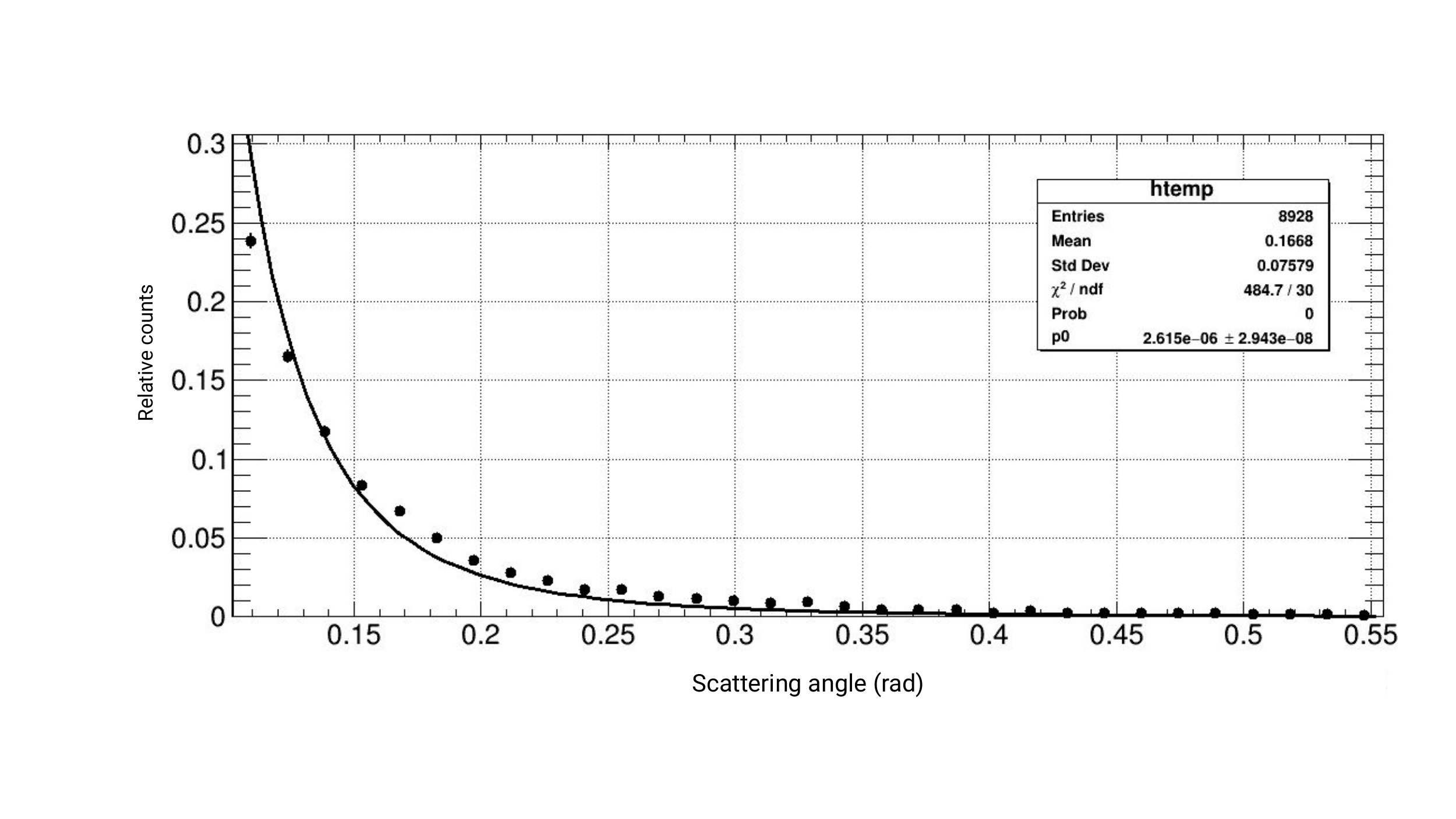}
    \caption{Fit of normalised distribution of the scattering angle to the function $p0\cdot cosec^4(\theta/2)$, where $p0$ is the fitting parameter. In the plotting tool, the `norm' option can be used to normalise histograms.  }
    \label{fig:alpha-fit}
\end{figure}

The first observation from the plot is the two low count bins for negative scattering angles,  which is an indication of backscattering. These will not be observed if the number of events is too low. In the original paper by Rutherford, it is seen that the backscattering was about 1 in 20,000, which corresponds to the numbers seen in this simulation \cite{rutherford2012scattering}. Next, in order to verify that the intensity is proportional to $cosec^4(\theta/2)$, the normalised histogram is fitten to this function, as shown in figure  \ref{fig:alpha-fit}. 

As can be seen, the distribution follows a $cosec^4(\theta/2)$ function with the fitting parameter estimated as $2.6\times10^{-6} $. This fitting parameter (shown as $p0$ in the figure) is equal to the prefactor in equation \ref{eqn:rcs3}, i.e.,
\begin{equation}
    p0 = \frac{d\Omega \cdot N_A \cdot \rho \cdot t}{A}  \left(\frac{ZZ'\alpha\hbar c}{4E}\right)^2
    \label{alpha-prefactor}
\end{equation}
From equation \ref{alpha-prefactor}, the term $\left(\frac{ZZ'\alpha\hbar c}{4E}\right)^2$ can be estimated from the knowledge of the fitting parameter and rest of the other terms $N_A$, $\rho$, $t$, $A$ and $d\Omega$.

In an experiment, $d\Omega$ is estimated from the detector opening angle and its distance from the source. In the simulation, we have a detector (sphere) that covers the full solid angle. However, in the histogram using which the fitting was done, binning in $\theta$ leads to a solid angle that varies as a function of $\theta$. The calculations are omitted here for the sake of brevity but it was seen that $d\Omega=0.1$ $sr$ yielded an estimate of  $\left(\frac{ZZ'\alpha\hbar c}{4E}\right)^2$ as 1.10 $\times$ $10^{-24}$ $cm^{2}$. 

The term $\left(\frac{ZZ'\alpha\hbar c}{4E}\right)^2$ can also be directly estimated  as all the parameters in this term are known. This direct estimate using the parameters given in table \ref{tab:alpha-param2} is $1.29\times10^{-24}$ $cm^2$. 

This simulation can also be extended to study the effects of beam attenuation as a function of target thickness. For more realistic estimation, the simulation can be performed using a geometry similar to the experimental setup used by Geiger and Marsden, in which the solid angle estimation is straight-forward \cite{gegier1909diffuse}. 

\begin{table}[]
\begin{tabular}{ll}
Parameter           & value               \\ \hline
$N_A$                & 6.023$\times$$10^{23}$ \\
$\rho$ & 19.32 $g/cm^3$                           \\
t                   & 0.0004 $cm$                              \\
A                   & 196.96657 $g/mol$                        \\
Z                   & 2                                       \\
Z'                  & 79                                        \\
$\hbar c$           & 197 MeV fm                         \\
E                   & 5 MeV                                   \\ 
$\alpha$              & 1/137 \\ \hline
\end{tabular}
\caption{Parameters used in the calculation of the prefactor in Rutherford's scattering formula.}
\label{tab:alpha-param2}
\end{table}

\subsection{Proton Bragg peak}
\begin{figure}[htpb]
    \centering
    \includegraphics[trim={4cm 6cm 3cm 4.0cm},clip,scale=0.75]{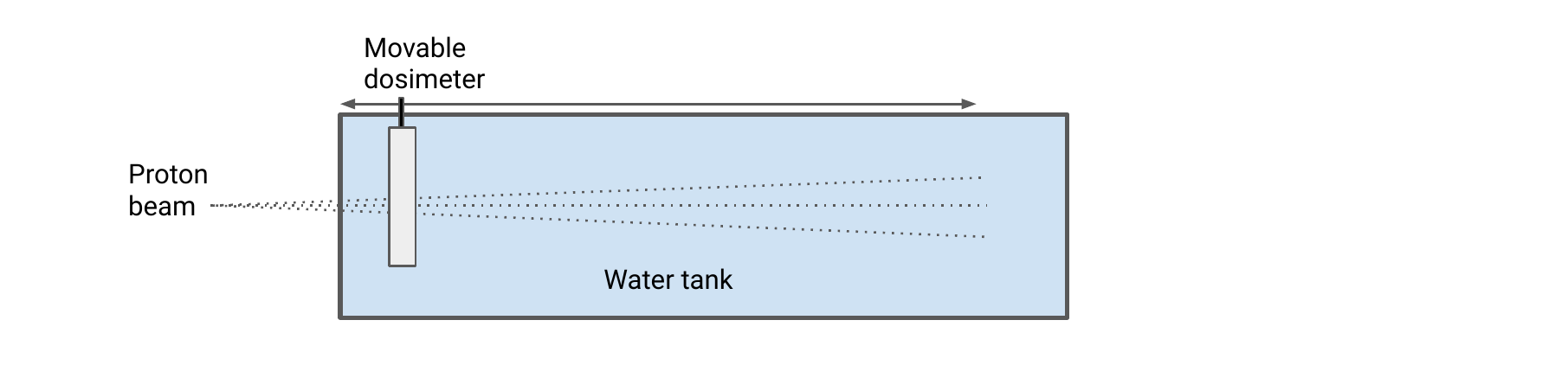}
    \caption{Typical setup used to measure the depth-dose distribution of protons. The dosimeter is moved along the beam axis and the dose is measured as a function of depth.}
    \label{fig:proton-setup}
\end{figure}

The use of scoring volumes can be demonstrated through the simulation of proton energy loss in water. The energy loss of protons in matter is defined by the Bethe-bloch equation \cite{paganetti2018proton}. The energy loss curve is characterised by a flat plateau followed by a peak and a fall-off due to the $1/v^2$ ($v$, velocity of the particle) nature closer to the end of the range. This feature is used in proton therapy for cancer treatments. The range of protons is therefore an important parameter for quality assurance routines in clinics.   Experimentally, the range of protons is measured in a water tank using a dosimeter. The dosimeter is moved along the axis of the beam and the dose is measured at every depth. The range of the beam is estimated from this depth-dose distribution, which also exhibits the Bragg peak. The range is defined to be the depth at which the dose falls to 80\% of the peak value at the distal end, clinically called as $d_{80}$. A schematic of a typical setup is shown in figure \ref{fig:proton-setup}.

The features of the proton depth-dose distribution along with the estimation of the range can be demonstrated in the following simulation.

\subsubsection{Simulation setup}
\begin{figure}[htpb]
    \centering
    \includegraphics[trim={5cm 1cm 8.5cm 0cm},clip,scale=0.70]{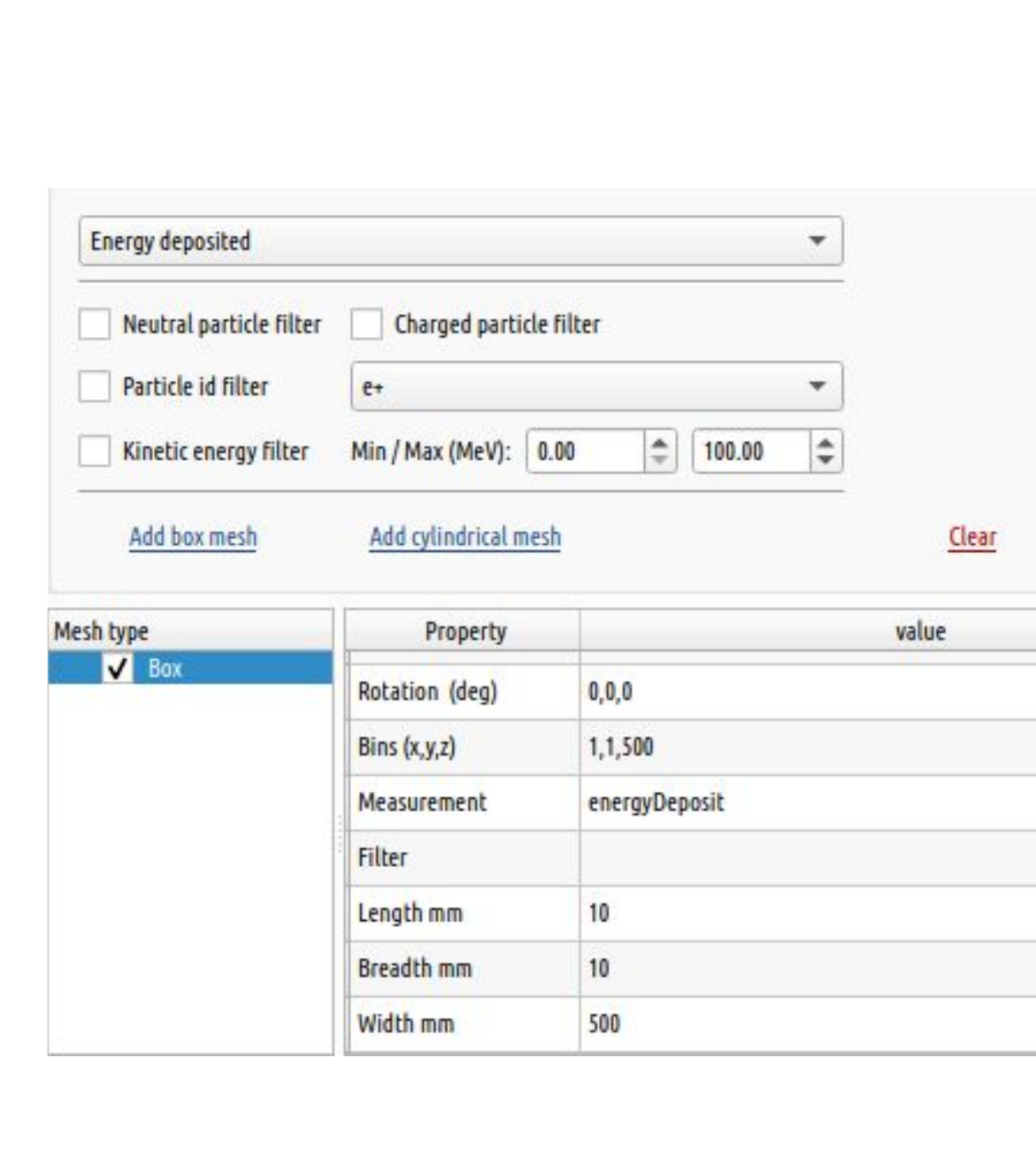}
    \caption{Screenshot of UI for creating a scoring volume in Simple.}
    \label{fig:scoringui}
\end{figure}

The FTFP-BERT physics list or the proton therapy physics list can be used for this simulation. A cube of dimension 10 mm $\times$ 10 mm $\times$ 500 mm with material set to water is kept inside a world volume (cube)  with similar dimensions but with an additional margin of 1 mm. Protons of 100 MeV energy are shot from one end of the water volume (along the $-z$ axis, in this case).
A scoring volume with the same dimension and position as the water volume is created with a binning of 1, 1, 500 along the $x$, $y$ and $z$ axis, respectively. This volume is set to measure the energy deposited. A screenshot of the UI for setting the scoring volume parameters is shown in figure \ref{fig:scoringui}.

\subsubsection{Data analysis}
\begin{figure}[htpb]
    \centering
    \includegraphics[trim={6cm 3cm 3cm 3cm},clip,scale=0.75]{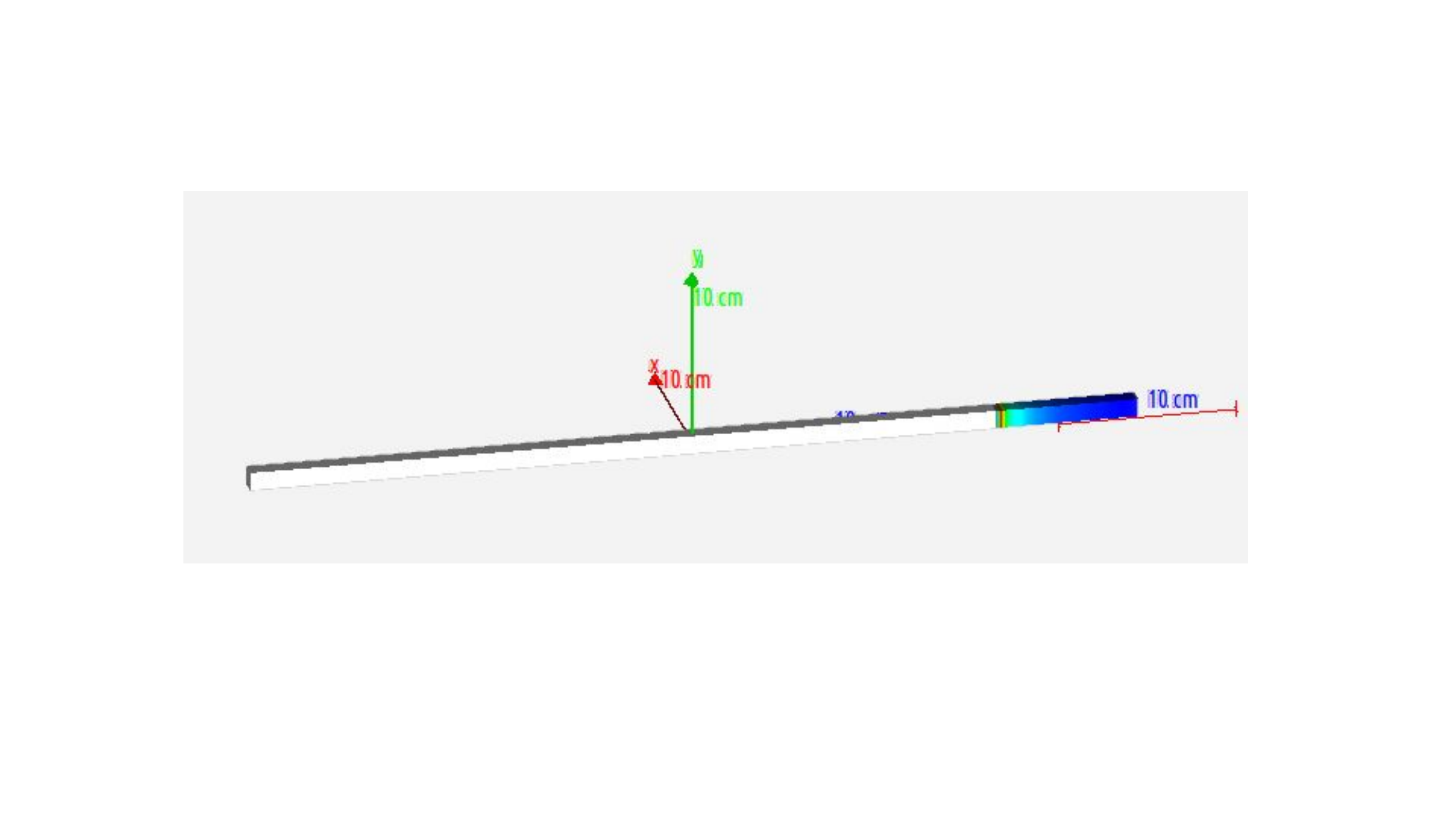}
    \caption{Projection of the dose distribution overlaid on the geometry.}
    \label{fig:proton-projection}
\end{figure}

For the simulation, about 50,000 events were shot. At the end of the simulation, the data from the scoring mesh is automatically populated in the dataframe. The dataframe contains the columns named \emph{binx}, \emph{biny}, \emph{binz}, \emph{var} and \emph{entries}. The \emph{var} column contains the energy deposited in the specific bin denoted by \emph{binx}, \emph{biny} and \emph{binz} and the \emph{entries} column gives the number of entries in that specific bin. The map of the scored quantity (i.e., energy deposited) can be overlaid on the geometry by choosing the `Plot projection' option from the context menu of the scoring volume. A screenshot of this projection is shown in figure \ref{fig:proton-projection}.

\begin{figure}[htpb]
    \centering
    \includegraphics[trim={3cm 1cm 0cm 0cm},clip,scale=0.45]{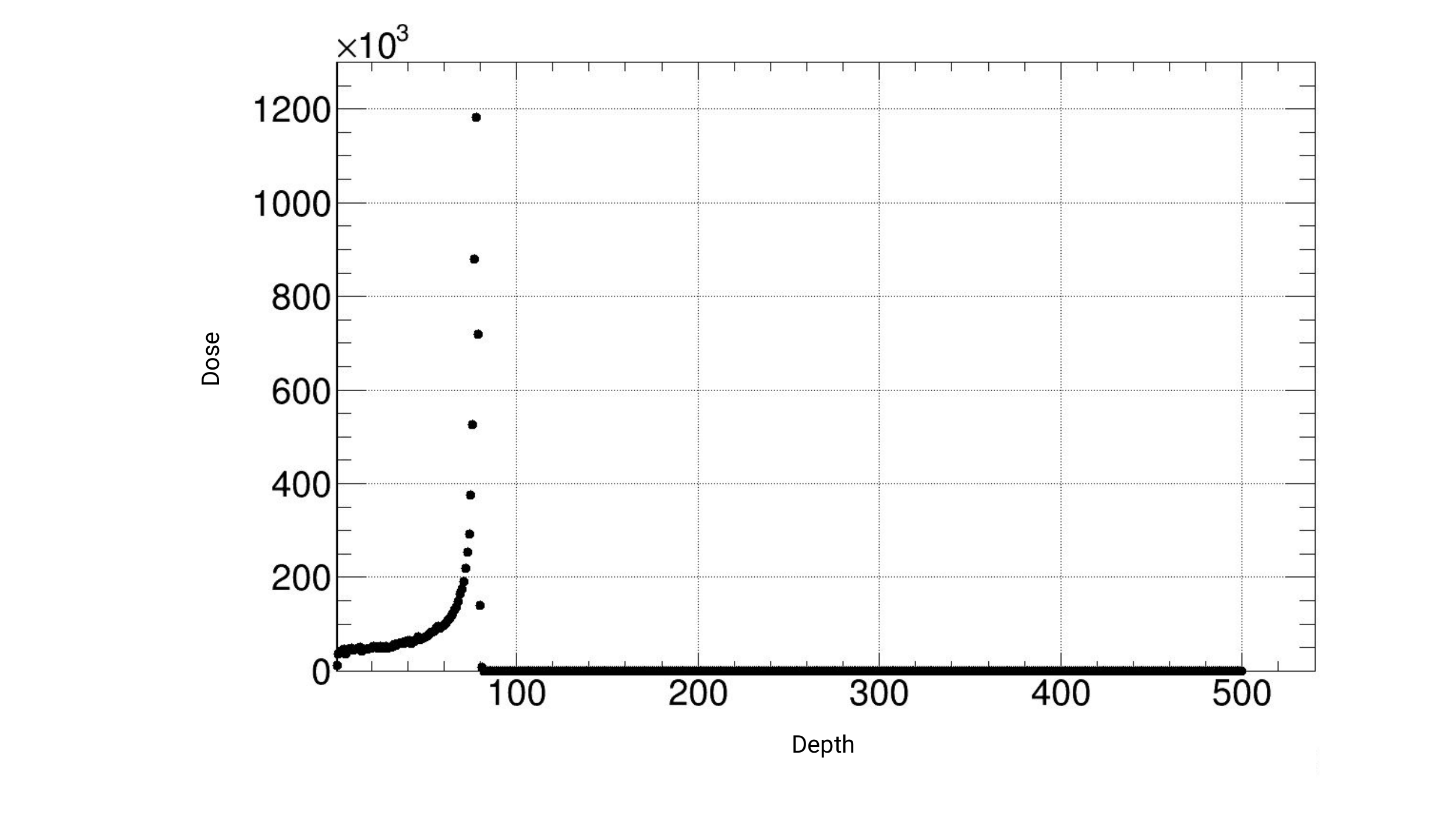}
    \caption{Depth-dose distribution from the simulation obtained by using the plot string \emph{var:binz} in Simple. The depth is in units of bins and the dose is in arbitrary units. The range is the depth at which the dose falls to 80\% of the peak dose at the distal (falling) part. A rough estimate of the range can be obtained by considering that the dose falls to zero at the falling edge at around 80 bins, which corresponds to 80 mm.}
    \label{fig:proton-bragg}
\end{figure}

The depth-dose distribution can be plotted by using the string \emph{var:binz} in the plot window. The graph is shown in figure \ref{fig:proton-bragg}.

The distribution shows the plateau region, followed by the Bragg peak and the distal end, beyond which the dose is zero. The range can be found by determining the depth at which the dose falls to 80\% of the peak dose at the distal end. This requires an extrapolation of the points at the falling edge. However, a rough estimate can be obtained from the graph by noting the first point at which the dose becomes zero. This value is about 80 bins, which corresponds to 80 mm, close to the NIST range of 77 mm for 100 MeV protons in water \cite{berger1998stopping}.

This simulation can be extended by studying the beam range as a function of energy and by changing the particle source to pions and other heavy ions.

\subsection{Particles in a magnetic field}
A final example to demonstrate the feature of magnetic fields in Simple is explained in this section. In many particle physics experiments, the momentum of a particle is estimated from its trajectory in a magnetic field. In this simulation, we design a magnetised sampling calorimeter using which one can study particle propagation in magnetic fields or use the data as an input for track reconstruction algorithms.

\subsubsection{Simulation setup}
\begin{figure}[htpb]
    \centering
    \includegraphics[trim={3cm 1cm 0cm 0cm},clip,scale=0.55]{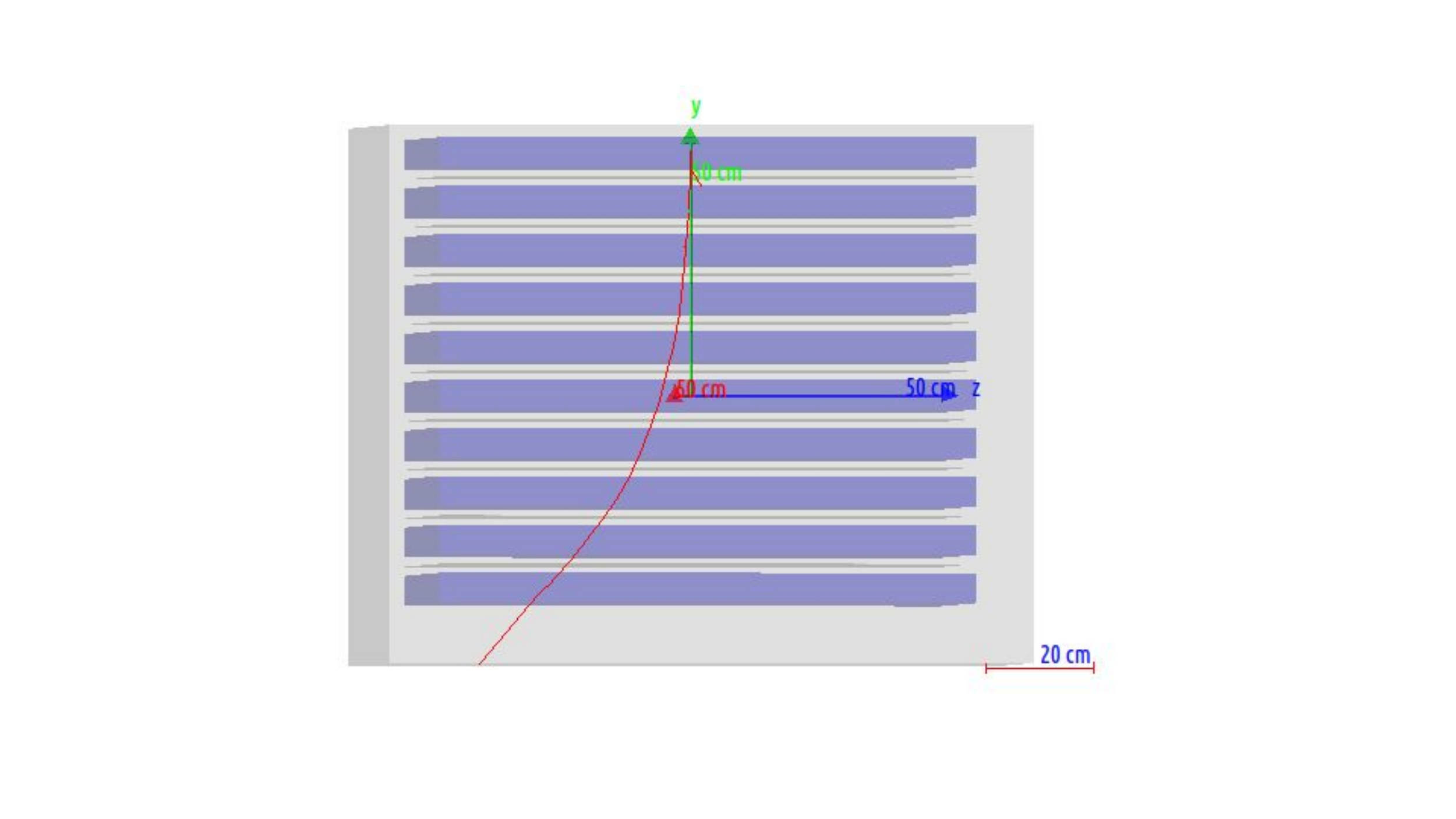}
    \caption{Screenshot of the simulation setup of a magnetised sampling calorimeter showing the track of a particle.}
    \label{fig:mical}
\end{figure}

The FTFP-BERT physics list is used for this simulation as well. The sampling calorimeter consists of 10 cubes, each of dimension 1000 mm $\times$ 1000 mm $\times$ 56 mm. The spacing between these volumes is kept at 180 mm. In cases where volumes are repeated, the `duplicate' feature, which can be accessed from the context menu of a volume, is used. The material is set to iron (G4\_Fe material in Geant4). The magnetic field in all these volumes is set to 15 kG along the $x$ axis. In between these volumes, a thin sheet (cube) of air of dimension 1000 mm $\times$ 1000 mm $\times$ 0.01 mm is placed. This sheet will act as a detector to track the position of the particles. This setup is placed inside a world volume (cube) of dimension 1200 mm $\times$ 1000 mm $\times$ 1200 mm and the material is set to air. 
Muons of various energies can be shot from the top and the trajectory of the particles can be viewed (and are also stored in an output file). A screenshot of the simulation setup along with a particle track is shown in figure \ref{fig:mical}.

A reconstruction algorithm can be written to estimate the momentum of the particles from the hits  stored in the output file. This simulation can be extended to study particle trajectory for different magnetic fields, particle energies, particle types and incident angles.

\section{Conclusions}
In this paper, we have presented Simple, a Geant4-based tool for simplifying particle physics simulations.  Simple eases the workflow of simulations in particle and nuclear physics and can be used as a tool for teaching many interesting concepts in a classroom setting. At the same time, Simple can also be used for advanced analyses as it uses the same simulation engine as Geant4. Simple is distributed as a standalone program that requires no installation or prior knowledge of Geant4. This feature can be helpful in tutoring beginners with no introduction to programming. Also, the in-built ROOT-based plotting tool facilitates basic analysis of the simulation results.

Simple is open-sourced and the code is available at \href{https://github.com/deepaksamuel/simple}{https://github.com/deepaksamuel/simple} with the standalone application generated for Ubuntu. In future, depending on user request, this application will be created for other operating systems including Windows 10. Advanced features like creation of boolean solids and inclusion of event generators are also planned. In addition, development of a module for creation of commonly used detectors like scintillation detectors, GM counters, resistive plate chambers, etc., is also a part of the future plan.

\section{Acknowledgements}

The author gratefully acknowledges the India-based Neutrino Observatory collaboration for the enriching research environment it provided in which many ideas, including Simple, were born. The author is thankful to the students of the Department of Physics, Central University of Karnataka, for their useful inputs in improving the tool. The author also thanks SD for the painstaking efforts in proofreading the manuscript.

\bibliographystyle{elsarticle-num}
\bibliography{mybibfile}

\end{document}